\documentclass[12pt]{article}
\usepackage{psfig,epsfig,cite,eqsection}

\newcommand{\<}{\langle}                                     
\renewcommand{\>}{\rangle}

\newcommand{\simg}{\stackrel{>}{\sim}}

\newcommand{\reff}[1]{(\ref{#1})}
\newcommand{\bea}{\begin{eqnarray}}
\newcommand{\eea}{\end{eqnarray}}
\newcommand{\be}{\begin{equation}}
\newcommand{\ee}{\end{equation}}
\newcommand{\beq}{\begin{equation}}
\newcommand{\eeq}{\end{equation}}

\newcommand{\la}{\langle}
\newcommand{\ra}{\rangle}
\newcommand{\li}{\left(}
\newcommand{\ri}{\right)}
\newcommand{\lc}{\left[}
\newcommand{\rc}{\right]}
\newcommand{\q}{{\bf q}}

\newcommand{\eps}{{\epsilon}}

\newcommand{\ome}{\mbox{\boldmath $\omega$} }
\newcommand{\x}{{\bf x}}
\newcommand{\e}{{\bf e}}

\newcommand \climit {\underset{R \to \infty}{\overset{n \to 0}
             {-\kern-0.5em - \kern-0.5em -\kern-0.5em \longrightarrow}} }

\begin{document}
\setlength{\unitlength}{0.2cm}

\title{A Simple Model for the DNA Denaturation Transition.}

\author{
  \\
  {\small Maria Serena Causo, Barbara Coluzzi, and Peter Grassberger} \\[-0.2cm]
  {\small\it John von Neumann-Institut f\"ur Computing (NIC)}  \\[-0.2cm]
  {\small\it Forschungszentrum J\"ulich}  \\[-0.2cm]
  {\small\it D-52425 J\"ulich, GERMANY}          \\[-0.2cm]
  {\small Internet: {\tt M.S.Causo@fz-juelich.de}}     \\[-0.2cm]
  {\small Internet: {\tt B.Coluzzi@fz-juelich.de}}     \\[-0.2cm]
  {\small Internet: {\tt P.Grassberger@fz-juelich.de}}     \\[-0.2cm]
  {\protect\makebox[5in]{\quad}}  
  \\
}
\vspace{0.5cm}

\maketitle
\thispagestyle{empty}   

\def\spose#1{\hbox to 0pt{#1\hss}}
\def\ltapprox{\mathrel{\spose{\lower 3pt\hbox{$\mathchar"218$}}
 \raise 2.0pt\hbox{$\mathchar"13C$}}}
\def\gtapprox{\mathrel{\spose{\lower 3pt\hbox{$\mathchar"218$}}
 \raise 2.0pt\hbox{$\mathchar"13E$}}}

\vspace{0.2cm}

\begin{abstract}
We study pairs of interacting self-avoiding walks $\{\ome^1,\ome^2\}$ 
on the 3d simple cubic 
lattice. They have a common origin $\ome_0^1 = \ome_0^2$, and are allowed 
to overlap only at the same monomer position along the chain: $\ome_i^1
\neq \ome_j^2$ for $i\neq j$, while $\ome_i^1=\ome_i^2$ is allowed. 
The latter overlaps are indeed favored by an energetic gain $\eps$. 

This is inspired by a model introduced long ago 
by Poland and Sheraga [J. Chem. Phys. {\bf 45}, 1464 (1966)] 
for the denaturation 
transition in DNA where, however, self avoidance was not fully taken into 
account. For both models, there exists a temperature $T_m$ above which 
the entropic advantage to open up overcomes the energy gained by forming 
tightly bound two-stranded structures.

Numerical simulations of our model indicate that the transition is of 
first order (the energy density is discontinuous), but the
analog of the surface tension 
vanishes and the scaling laws near the transition point are exactly 
those of a second order transition with crossover exponent $\phi=1$.
Numerical and exact analytic results show that the transition is 
second order in modified models 
where the self-avoidance is partially or completely neglected.
\end{abstract}

\pagebreak

\section{Introduction}
The study of the nature of the DNA denaturation is a long standing
open problem.
Experimentally a multistep behaviour in light absorption
as a function of the temperature was observed already in the
fifties (see \cite{review} as a review).
This suggested a sudden sharp opening of clusters of base pairs
in cooperatively melting regions. This scenario is reminiscent of 
the behaviour at a discontinuous  first order phase
transition, in which the system changes its state {from} a double
strand to two molten single-stranded chains. Since then, this 
scenario has been verified and studied in great detail \cite{review}.

Early theoretical attempts to model this transition could not reproduce
these phenomena. The first attempt with a one-dimensional 
Ising model in which the two states
of spin correspond to a open or close state of the base pair,
with a favorable coupling between neighbor pairs that are in
the same state \cite{ZiBr1,ZiBr2}, reproduced a crossover
between the two different regimes but no thermodynamical transition.

The first refinement consisted in taking into account the different
entropic weights of opened bubbles and double stranded segments
\cite{PoSh}, since the phase space region that two terminally 
joined (but otherwise free) open strands can explore is bigger than 
the one accessible to a double strand of the same length.

This model was solved using the entropic weights of
self-avoiding loops in refs.\cite{Fi,Fisher_83}. In 
this way the self avoidance between bases within the same loop 
is taken into account, 
but the other mutual excluded volume effects are completely neglected. 
This simplified model displayed a smooth second order 
transition in two and three dimensions.

These models were of course only very rough caricatures of the 
true complexity of the problem. Even if we believe that microscopic 
details should be irrelevant for the existence and order of the 
DNA melting transitions, there are a number of aspects which one 
might suspect to be relevant. In addition to self avoidance these 
include the stiffness of DNA,
the difference in stiffness between single- and double-stranded 
DNA, the different properties of A-T and C-G pairs, and the helical 
structure of double-stranded DNA. Finally, one should also consider 
the effect of ``wrong" base pairings, either between bases of the 
two different strands or between bases within the same strand.

There seems to exist up to now no model which incorporates all these 
aspects. But there have been recent models where some of them where 
included, and which seem to reproduce the sudden opening of base pairs.
The common property in all of them is an entropic barrier
that favors configurations in which base pairs are far apart.

The `nonlinear model', introduced in \cite{DaPe1,DaPe2}, assumes that
the stacking energy between neighboring base pairs depends on whether
these pairs are in `helical' or `coil' states (i.e., whether they are
bound in a double string or not). In a helix, this stacking energy is
larger than in a coil.
Transfer-integral calculations, molecular dynamics simulations \cite{DaPe1}
and approximate analytical methods \cite{DaPe2} pointed out a 
first order phase transition.

In the same direction goes also a recent model \cite{MariaSimona} in which
the helical structure is taken seriously. As a result, 
a mechanical torque 
which tends to increase or decrease the winding becomes a new 
thermodynamical variable. 
A transfer matrix calculation \cite{Co} shows that this model
exhibits a first order phase transition in the
temperature - torque plain, analogous to the liquid-gas
transition in the temperature - pressure plane.

Finally, according to a recent study \cite{CuHwa}, the effect of
the heterogeneity in the DNA sequence -- which amounts to a frozen 
disorder in the base pair binding strength -- has no effect on the 
order of the transition 
if the model contains no entropic barrier. But it gives rise to a
multistep energetic landscape if a state dependent stiffness of the type
considered in \cite{DaPe1} is introduced.

In the present paper we consider a simplified model where all these 
features are disregarded, but -- in contrast to the papers 
mentioned above -- excluded volume interactions are fully incorporated.
Our model consists of two interacting 
self-avoiding walks, corresponding to the two single strands, with the 
same origin on a 3d cubic lattice. Each monomer 
corresponds to a base and is supposed to have its complementary 
at the same position in the other chain. Two monomers with 
different positions in the two chains are not allowed to occupy the 
same lattice site, whereas the overlap of monomers at the same
position is favored by an energetic gain $\eps$ that represents
the binding energy. Base-pair misalignments are forbidden.
We consider the homogeneous case, where all the binding energies are equal.

In our approach we focus mainly on the two conflicting tendencies of 
the system: the entropic gain due to the larger number of
configurations accessible to the two open strands system
on one hand and the tendency to build energetically favored links between 
the two strands on the other. The necessity to balance these opposite 
tendencies when minimizing the free energy leads to 
the finite-$T$ phase transition between the high
temperature swollen phase, and the low temperature phase in which finite 
fractions of the chains overlap.

\section{The model}

Let us define two $N$-step chains with the same origin on 
the 3-$d$ lattice $\ome^1 = \{\ome^1_0, \dots, \ome^1_N\}$ and 
$\ome^2 =\{\ome^2_0, \dots, \ome^2_N\}$ with $\ome^k_i \in Z^3$ 
and $\ome^1_0=\ome^2_0 = (0,0,0)$.

The Hamiltonian (or rather Boltzmann weight) which describes 
a configuration $(\ome^1,\ome^2)$ of our system is
\be
   e^{-H \over KT} = \prod_{i\ne j}(1-\delta_{\ome^1_i,\ome^2_j})
(1-\delta_{\ome^1_i,\ome^1_j})(1-\delta_{\ome^2_i,\ome^2_j})
\exp\left({-{\hat \eps}\over kT}\sum_{i=0}^N \delta_{\ome^1_i,\ome^2_1}\right) .
\ee
Thermodynamic properties of the system only depend on the
reduced variable $\eps=-{\hat \eps}/KT$ that we will use in the
following.
The partition sum can therefore be written as 
\be
   Z_N(\eps) = \sum_{n=0}^{N} c_{N,n} e^{\eps \:n}
\ee
where $n$ is the number of contacts, $n={\rm card}\{i | \ome^1_i=\ome^2_i,
i>0\}$, and $c_{N,n}$ is the number of distinct configurations with 
$n$ contacts (notice that $\langle n\rangle/N$ is the natural order 
parameter). Alternatively, by introducing a fugacity $z$ we can go 
over to the grand canonical ensemble with partition sum
\be
   G(z,\eps) = \sum_{N=0}^\infty z^N Z_N(\eps) = \sum_{N=0}^\infty
                   \sum_{m=0}^{N} c_{N,n} e^{n\eps} \;.
\ee

While we fix the starting point of the two polymers at a same origin,
$\ome^1_0 = \ome^2_0$, we allow the end-points to wander freely 
in space. This is different {from} the Poland-Sheraga model 
\cite{PoSh}, where also the end-points were forced to coincide, 
$\ome^1_N= \ome^2_N$. 
At least in the ideal case, 
the presence of this constrain does not affect the order of the transition and this
should be true also when the excluded volume interaction is taken into account.
However the crossover scaling functions  between different regimes are not the same in the two models 
and at the tricritical point, at which the transition takes place,
 different entropic exponents are found.
In the excluded volume case there is also a topological subtlety: 
if chains are deformed continuously, 
non-trivial knots are forbidden if the end points never separate. In 
contrast, in our thermodynamical treatment any knots are allowed. But 
this should not have much influence either.

\section{Approximate treatment}

The system can be represented as a sequence of $M$ superimposed 
self-avoiding walks of length $n_1, \dots, n_M$ ($n_k\geq 0$ with $n=\sum_k(n_k+1)-1$),  
which correspond to helical domains in DNA where base pairs are bound together,
 which alternate with $M-1$ bubbles of lengths $p_1, \dots, p_{M-1}\; (p_k\geq 1;$ 
molten regions). On the lattice, they are self-avoiding polygons of length
$2 p_i$. The last part consists of two self-avoiding walks of lengths 
\be
   r\li \{n,p\} \ri = N- \sum_{i=1}^M n_i - \sum_{j=1}^{M-1} p_j . 
\ee
All the elements of the sequence must be mutually avoiding, that means that two monomers
can occupy the same position in the space only if they occupy the same position along
the chain. 

If this last constraint is neglected, one can factorize the problem
and write a generating function for the system in terms of the
generating functions of single self-avoiding walks, polygons,
and a pair of self-avoiding walks starting at the same origin.
This leads to the `almost unidimensional' phase transitions 
of \cite{PoSh,Fi}. 

The partition sum in the fixed-$N$ ensemble can be written as
\be
   Z_N = \sum_{\{n,p\}} S_{r(\{n,p\})} \prod_i W_{n_i} V_{p_i} 
\ee
where the sum runs over all possible partitions into helices and 
bubbles. $W_n=e^{\eps n/{kT}}c_n$, $V_p={\cal C}_{2p}$ and
$S_{r\li \{n,p\} \ri}=c_{2r\li \{n,p\} \ri}$. 
Here $c_n$ is the number of self-avoiding walks of length 
$n$, while ${\cal C}_{2p}$ is 
the number of self avoiding polygons of length $2p$.
This partition sum is clearly an upper bound to the true one, since
many configurations are included which are not allowed due to self 
avoidance.

The simplified problem can be easily solved in the grand canonical
ensemble, i.e. by considering the generating function 
$G(z)=\sum_{N=0}^\infty z^N Z_N$. We find
\be
   G(z) = \frac{G^\eps_W(z) G_S(z)}{1- e^\eps \: G^\eps_W(z) G_V(z)}
\label{generatingfunction}
\ee
where 
$G^\eps_W(z) = \sum_{N=0}^\infty z^N W_N$, $G_V(z) = \sum_{N=0}^\infty z^N V_N$, 
$G_S(z) = \sum_{N=0}^\infty z^N S_{N}$.
The critical behaviour of the system is determined by the singularity
of $G(z)$ which is closest to the origin, and it can be studied by
using the asymptotic forms for the number of self-avoiding walks and polygons
$c_N \approx \mu^N N^{\gamma -1}$,
${\cal C}_N \approx \mu^N N^{\alpha-2}$, where $\mu$ is the connectivity constant, 
which is a lattice dependent quantity. 
Two cases are possible: that the singularity closest to the origin comes from
$G_S(z)$ or from vanishing of the denominator.
Let us focus on the case $d=3$.
Recent estimates of the critical exponents are
$\alpha= 2 - d \nu = 0.23723(4)$ (where we used the estimate $\nu=0.58758(7)$ 
given in \cite{Nickel}) and  $\gamma = 1.1575 \pm 0.0006$ \cite{Serena}.

Let us denote with $z_W$, $z_V$ and $z_S$ the location of the
singularities of the generating functions on the real axis. Since
\be
G^\eps_W(z) = \sum_{N=0}^\infty \frac{\left(ze^\eps \mu\right)^N}{N^{1-\gamma}} \;\;\;\;
G_V(z) = \sum_{N=0}^\infty \frac{\left(z\mu^2\right)^N}{(2N)^{2-\alpha}} \;\;\;\;
G_S(z) = \sum_{N=0}^\infty \frac{\left(z\mu^2\right)^N}{(2N)^{1-\gamma}}
\ee
it appears immediately that $z_W=\frac{e^{-\eps}}{\mu}$, 
$z_S=\frac{1}{\mu^2}$, while $G_V(z)$ is finite at
$z_V=z_S=\frac{1}{\mu^2}$ (since one has $2-\alpha > 1$), but it diverges 
for $z > z_V$.

In the high temperature regime, i.e. $\eps \to 0$, the singularity
of $G_S$ is the first to occur and the critical behaviour is that of
two self-avoiding walks. This means that the system is in the denaturated
state, and the corresponding free energy density is given by 
$f/KT=\log z_S=-2 \log \mu$. This is just the entropy
density of two self avoiding walks with the same origin.

Since $G^\eps_W(z)$ is an increasing function of $\eps$,
for decreasing temperatures (increasing $\eps$) the
zero $z_{WV}$ of the denominator in Eq.\reff{generatingfunction} decreases 
and finally becomes lower than $z_S$. The crossing point
corresponds to the melting transition $\eps^*$.

It can also be shown that the order of the
transition is determinated {from} the singular behaviour of $G_V(z)$ in $\frac{1}{\mu^2}$, namely on
the value of the exponent $2-\alpha$ \cite{Fi}. 
Since $G^\eps_W(z)$ is regular in $\frac{1}{\mu^2}$ at $\eps=\eps^*$, 
it plays an irrelevant role at the transition point and this is independent
from the value of $\gamma$, i.e. the fact that the helical domanis are self-avoiding
or ideal does not affect the order of the transition.
The free energy for $\eps \ge \eps^*$ is given by
\begin{equation}
f/KT = \left \{ \begin{array}{lcl} 
-2 \log \mu + C (\eps-\eps^*)^{1/(1-\alpha)} + \dots & \mbox{ for } &  1 < 2-\alpha < 2 \\
-2 \log \mu + C (\eps-\eps^*) + \dots & \mbox{ for } & 2-\alpha > 2
\end{array} \right. \end{equation}

Since $2-\alpha =1.76276(6)$ \cite{Nickel}, the approximate solution predicts
a second order phase transition.

The main approximation involved in the above treatment is that it neglects 
excluded volume effects which come {from} the mutual interactions of 
different bubbles, segments and free terminations the one with the others.
As already pointed out, this means that we overestimate the
partition function, and the transition could be sharper than predicted by 
this simple model.

On the other hand, one can immediately use the above arguments 
to infer that the transition is certainly of second order in the 
case of interacting random walks, since $2-\alpha=3/2$ there, 
and there are no excluded volume effects to be taken into account. 
We present in the appendix an exact analytical treatment of
the ideal case. There we also evaluate numerically the melting value $\eps^*$, 
and we study the scaling laws whose general structure will be discussed 
in the next section.

\section{Scaling laws}
\label{scl}

Again it is more easy to discuss the problem in the grand canonical 
ensemble, with the fugacity $z$ conjugate to $N$. The limit $N \to \infty$ 
in the monodisperse ensemble corresponds to $z \nearrow z_c(\eps)$. 

\begin{figure}[htb]
\vglue -.1cm
\centerline{\epsfig{file=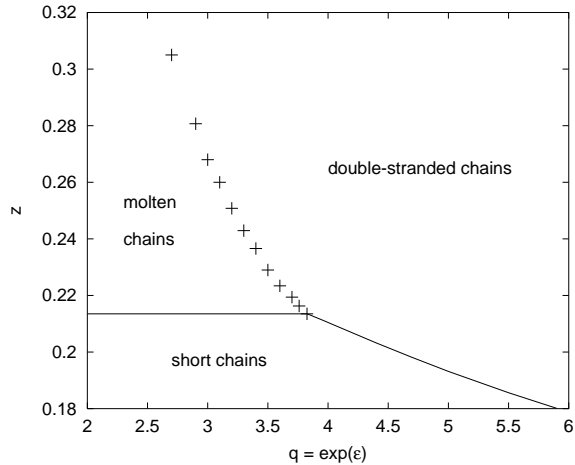,angle=270,width=8cm}}
\caption{Phase diagram. On the horizontal axis is plotted the Boltzmann factor $q=e^\eps$
  per bound monomer pair, while the fugacity is plotted vertically. Below the 
  continuous line, chains are short with an essentially exponential distribution 
  in chain length. At this line, the average chain length diverges. To the left 
  of the triple (and tricritical!) point, the line is horizontal (i.e., the 
  critical fugacity is independent of $q$ and coincides with the value for 
  normal SAWs). The ``molten chains" and ``double-stranded chains" phases are 
  well defined only for finite volume $V$, with the chain length $N\propto V$.
  The numerical determination of the phase boundaries is discussed 
  in Sec. 6. While the continuous line is very precise (error less than 
  the width of the line), the uncertainty of the molten/double-stranded phase 
  boundary is at least as big as the symbol size.}
\label{phase-fig}
\vglue -3mm
\end{figure}

Values 
$z>z_c(\eps)$ make only sense after placing the system in a (large but) 
finite volume $V$. The two polymers are allowed to grow until they fill 
the volume with a finite non-zero density $\rho = 2\langle N\rangle/V$ 
which remains constant in the limit $V\to\infty$. In this regime, we 
actually have two different phases, corresponding respectively to molten 
and undenatured (double-stranded) chains. It is intuitively clear that 
a non-zero density favors the presence of contacts, because contacts 
in our model reduce the volume occupied by the monomer pairs. 
Therefore one expects that the transition point 
between the molten and double-stranded phases 
takes place at a lower value of the interaction parameter $\eps_L^{\rm dense}(z)$ 
when the fugacity $z$ increases.
%
This phase diagram is shown in Fig.~\ref{phase-fig}. 

Notice that the boundary between the 
molten dense phase and the short chain phase is strictly horizontal, as the 
attractive interaction plays no role along this transition line. 
Qualitatively, this phase diagram is very similar to that for a
polymer attached to an adsorbing surface 
\cite{eisenriegler,hegger} and to surface transitions in magnetic systems
\cite{diehl}. But in contrast to the latter, the boundary between the two
dense phases is not horizontal.

Using the $(z,\eps)$ representation, it is clear that the melting transition 
is a tricritical point. Its analogue in magnetic systems with surfaces is the 
special point \cite{diehl}. The curve $z=z_c(\eps)$ consists of two parts, a 
horizontal one for $\eps<\eps^*$ and a tilted one for $\eps>\eps^*$. 
At the melting point $(z^*,\eps^*)$, one sees a change of critical behaviour.
At finite large $N$, for $\eps < \eps^*$ the critical behaviour of a SAW of length $2N$
is observed, while for $\eps > \eps^*$ the system displays a double stranded behaviour.
At fixed $N$, for $\eps$ near $\eps^*$, 
a crossover between the tricritical behaviour and the double stranded one
($\eps > \eps^*$) or between the tricritical behaviour and 
the $2N$-SAWs one ($\eps < \eps^*$) is observed as $\eps$ tends to $\eps^*$.
The width of the crossover region decreases as $N$ increases.

In the following we shall discuss the scaling laws that would be expected if 
the melting transition is second order. If it is first order, it seems
at first not clear whether the usual scaling scenario (which is based
on the existence of a divergent length scale) still holds. We shall
see in Sec.\ref{results} that it does hold even then.
An analytic study in the ideal case (no excluded volume)
is shown in the Appendix.

Near a tricritical point, the partition sum is expected to scale as 
\be
   G(z,\eps) = (z^*-z)^{-\gamma^*} F \li (\eps-\eps^*) /(z^*-z)^\phi \ri\;, 
                            \label{scaling-ansatz}
\ee
with $\phi$ being called the crossover exponent.
The scaling function $F(x)$ is non-singular at $x=0$, from which follows 
\be
   Z_N(\eps^*) \sim (1/z^*)^N N^{\gamma^*-1}
                \label{Z_tricrit}
\ee
for the scaling exactly at the melting point\footnote{The value of $\gamma^*$
in the ideal case is computed in the appendix. We find $\gamma^* = 1 + \phi$ (see Eq.\reff{tr_id}).
On the other hand if both the extremities are bound together as in the Poland-Sheraga model, it is
easy to see that the crossover exponent $\phi$ does not change, but the absence of $G_S(z)$
in the numerator of Eq.\reff{generatingfunction} gives a different singular behaviour at
$\eps^*$. It is simple to see, following the same lines as in the appendix, that this gives
$\gamma^* = \phi$.}.

For $\eps > \eps^*$ and $z \to z_c(\eps)$ from below, $G(z,\eps)$ must scale as the partition 
sum for a (double-stranded) SAW, $G \sim {\rm const}/(z_c(\eps)-z )^\gamma$.
Therefore $F(x)$ must have a singularity at some finite value $x_0$ where it diverges as
\be
F(x) \sim {{\rm const}\over (x_0-x)^\gamma},
\ee
and for small $\eps - \eps^*$
\be
z^*-z_c(\eps) \sim x_0^{-1/\phi}(\eps - \eps^*)^{1/\phi};.        \label{phase-bdry}
\ee
Thus the crossover exponent $\phi$ describes how the critical fugacity depends
on the contact energy in the bound (non-molten) phase.

Finally, for $\eps<\eps^*$ and $z\nearrow z^*$ we must have 
$G \sim {\rm const}/(z^*-z)^\gamma$. This is the case if 
\be
    F(x) \sim (-x)^\sigma \quad,  \qquad x\to -\infty
\ee
with some power $\sigma$, and
\be
   \gamma^*+\phi\sigma = \gamma .
\label{sigma}
\ee

Performing the Laplace transform one checks easily that eq.\reff{scaling-ansatz} 
is obtained with the ansatz 
\be
   Z_N(\eps) = \mu(\eps)^N N^{\gamma^*-1} \Psi\li (\eps-\eps^*) N^\phi\ri
                                        \label{Z-scaling}
\ee
with $\mu(\eps) = 1/z_c(\eps)$. In order to get the right asymptotics for 
$\eps\neq \eps^*$, the scaling function $\Psi(x)$ -- which is related to $F(x)$ 
by a Laplace transform -- must scale as $|x|^{(\gamma-\gamma^*)/\phi}$ for 
$x\to \pm\infty$.

The scaling of the energy is obtained by deriving $Z_N$ with respect to $\eps$.
It is \cite{eisenriegler}
\be
   E_N(\eps) \sim \cases{ (\eps-\eps^*)^{1/\phi-1}N & for $\eps > \eps^*$ \cr
                          N^\phi                    & for $\eps = \eps^*$ \cr
                          1/(\eps^*-\eps)           & for $\eps < \eps^*$ } \;.
\ee
{From} this we see that a first order transition is obtained for $\phi=1$. The 
scaling of the specific heat is obtained by deriving once more with respect to 
$\eps$. One finds that the peak of the specific heat scales as $N^{2\phi-1}$ 
and is located at $\eps^*+ const/N^\phi$.

One can also look at the system from an extended scaling point of view 
\cite{Pfeuty,dosSantos,Derrida}. We define two correlation lengths $\xi_1$ 
and $\xi_2$. The second, which we call the {\it geometrical} correlation length,
is identified with the Flory radius of any of the two 
polymers, $\xi_2 = \<(\ome_N^1-\ome_0^1)^2\>^{1/2} = \<(\ome_N^2-\ome_0^2)^2\>^{1/2}$.
It follows the scaling law $\xi_2 \sim N^\nu$ in any phase.
The first, $\xi_1$, is the {\it thermal} correlation length. It is defined  as
the mean diameter of the molten `bubbles'.
In the bound phase we expect it to scale with $N$ and $(\eps-\eps^*)$
in the same way as the end-to-end distance
between the two strands, $\xi_1 \propto R_{\rm end}=\<(\ome_N^2-\ome_N^1)^2\>^{1/2}$ .
If the denaturation transition is second order 
the thermal correlation lenght $\xi_1$ inside the bound phase converges in the limit $N \to \infty$ 
to a constant which depends on $\eps$,
but the value of the constant diverges as $\eps \searrow \eps^*$.
Exactly at $\eps =\eps^*$ it should  scale as a function of $N$ in the same way as $\xi_2$ i.e. both 
correlation lengths should be equivalent. 

In contrast, in a usual first order transition we would 
expect that the thermal correlation lenght remains finite as $N$ tends to infinity
also in the limit $\eps \searrow \eps^*$, but we will see that in our system this picture does not hold 
because of vanishing of a surface tension.

In approaching the transition point from 
the molten phase $R_{\rm end}$ scales as the Flory radius and is then not related to $\xi_1$.

Denoting with a subscript $T$ the exponents that govern 
the scaling laws in the thermal parameter $\eps-\eps^*$,
we can define a thermal correlation length exponent $\nu_T$ by assuming 
$\xi_1 \sim (\eps^*-\eps)^{-\nu_T}$ in the limit where we take first $N\to\infty$
and then $\eps\to \eps^*-0$. One has \cite{dosSantos,Derrida}
\be
   \phi= \nu/\nu_T\;.                  \label{phi_nu}
\ee
This can be understood in two ways. First, one can invoke the fact that $\xi_2\sim\xi_1$
when $\eps=\eps^*$. Then Eq.\reff{phi_nu} expresses just the fact that $\partial \xi^2
/\partial\eps$ and $\partial \xi^2/\partial z$ are related by Eq.\reff{phase-bdry}.

Alternatively, one observes that $D=1/\nu$ is just the (Hausdorff-) dimension of 
the system, whence the specific heat exponent $\alpha_T = 2-\frac{1}{\phi}$ takes the familiar 
hyperscaling form $\alpha_T= 2-D\nu_T$ \cite{Derrida}.

Let us finally discuss histogram methods which have become increasingly popular 
during the last years. They provide expectation values at temperatures different 
from those used in the simulations. In addition, they are used to study finite 
lattice size effects \cite{wilding1,wilding2}. 
Near an ordinary temperature-driven critical point, the energy
distribution scales in a finite spin system of length $L$ as 
\be 
   P_L(E) \sim L^{-1/\nu} p\li(E-\<E_c\>)/L^{1/\nu}\ri 
\ee
This is different for a first order transition where the distribution has two 
peaks which get increasingly separated when the system size is increased. The 
minimum between the two peaks becomes exponentially deep (with the depth 
controlled by the surface tension between the two phases), and the
peaks become arbitrarily sharp in the limit $L\to\infty$. 

Instead of studying the distribution for fixed {\em lattice} size, in the 
present case it is natural to study it for fixed finite $N$. 
In view of $E\sim N^\phi$, one might now expect a similar behaviour 
\be
   P_N(E) \sim N^{-\phi} p(E/N^\phi) \;.           \label{PE}
\ee
In Sec. \ref{results} we show that this is indeed true for the melting transition 
of ordinary random walks in dimensions $2<d<4$. There the transition is second 
order with $\phi=d/2 -1$, and Eq.(\ref{PE}) is correct. But surprisingly Eq.(\ref{PE})
is also correct for ordinary random walks in dimensions $d>4$ where $\phi=1$ and 
the transition is first order. This can be understood as a vanishing of the analog 
of the surface tension: the cost involved in going over from a molten domain to 
a bound domain does not increase with $N$. This is obviously due to the fact that 
our system is (at least topologically) one-dimensional. Indeed we will show in 
Sec.\ref{results} that the same is also true for SAW melting in $d=3$.

\section{Simulations}

We use the pruned-enriched Rosenbluth method (PERM) \cite{Gr}, with
markovian anticipation \cite{FrCaGr}, which is particularly effective
to simulate interacting polymers \cite{se}. In the present case the
algorithm was implemented in such a way that the two chains grow 
simultaneously (i.e. adding one monomer to the first chain, than 
to the other, again to the first and so on\footnote{This can straightforwardly 
be extended to more than two chains and allows then very efficient 
simulations of star polymers, in particular near to collapse transitions
where other methods become inefficient.}). Following the PERM
strategy, the whole system grows according to the Rosenbluth  
method \cite{Ro} while  configurations with 
very large/very small weight are cloned/pruned. 
The bias used during the Rosenbluth sampling is corrected by multiplying the
weights of the configurations with the appropriate factor.

The $k$-steps markovian anticipation consists in an additional bias
based on the statistics of sequences of $k+1$ successive steps \cite{FrCaGr}. 
In dimension $d$, labeling by $s=0, \dots 2d-1$ the $2d$ directions on a hypercubic 
lattice and by $S=(s_{-k},\dots s_0)=({\bf s},s_0)$ a given sequence 
of $k$ steps ending in $s_0$, one considers the statistical weight 
$P_{N,m}(S)$ of all $N$-step chains which followed the sequence $S$ 
during the steps $N-m-k, \dots, N-m$.  
The bias in a $k$-step markovian
anticipation is then given by
\be
p(s_0|{\bf s})=P_{N,m}({\bf s},s_0)/ \sum_{s'_0=0}^{2d-1} 
P_{N,m}({\bf s},s'_0).
\ee
This means that a step in the direction $s_0$ is chosen more often
if the previous experience tells that it will be more successful in the
far ($m$ steps ahead) future. These biases are obviously compensated by
a factor $\propto 1/p(s_o|{\bf s})$, to get a correct sampling. 
In our simulation the weights $P_{N,m}(S)$ were estimated in a preliminary run. 
Moreover we used an  $ad~hoc$ bias for the present model. 
When the second chain has to perform a growth step and the end of the first one
is in a neighboring site, instead of doing a blind step multiplying then
the weight by a factor $e^\eps$ if the new contact is formed, we
favor it choosing the step towards the end of the first chain with
probability $ \propto e^\eps$.

One can further increase the probability of sampling configurations 
with many contacts by favoring growth steps which reduce the end
to end distance $r = |\ome_i^2-\ome_i^1|$. We found that such a bias 
(which has to go to zero for $r\to\infty$) can substantially enhance the 
sampling efficiency, but leads occasionally to ``glitches" where a 
disfavored configuration is encountered nevertheless, with exceptionally 
large weight. 
We therefore use only the previously described bias.

As one could expect for a first order transition, fluctuations near 
$\eps^*$ are $very$ strong, particularly on the probability
distribution of the energy. A large part of our statistics was collected
in order to obtain clear data for $P(E)$ up to large chain lengths 
($N=3000$, i.e. a total of $6000$ monomers), and for a very wide range 
of $E$. This aim was achieved by performing independent runs at
several different values of the interaction strength and by reweighting
results. The errors were evaluated with the jackknife
method, i.e. by using the fluctuations between independent runs. Since we 
have only few such runs to compare with, the errors should be considered 
just as rough estimates.

Moreover we made runs up to very large chains lengths ($N=8000$)
in order to study the large $N$ behaviour of the partition function
and of the end to end distance. We also performed simulations at finite
density (which will be discussed more in detail in the following) in order 
to locate the molten/doublestranded phase boundary.

As a last remark, we note that also in the ideal case in 5 dimensions, 
where it is much more simple to get good statistics, the use of a
bias is necessary for correctly sampling the energy distribution in a reasonable CPU time. 

\section{Results and discussion}
\label{results}

\subsection{The ideal case}
\label{id}

Before discussing self avoiding walks, we study first the case of ideal 
random walks, but of course weighting configurations with $n$ contacts 
by a factor $\exp(\eps \: n)$. 
The study of the ideal system, which is analytically solved in
the Appendix, allows us both to test the efficiency of the numerical
methods and to verify the peculiar first order transition predicted 
in 5 dimensions. We limit the analysis to simple hypercubic lattices 
with $d=3$ and 5. 

\begin{figure}[htb]
\vglue -1mm
\centerline{\epsfig{file=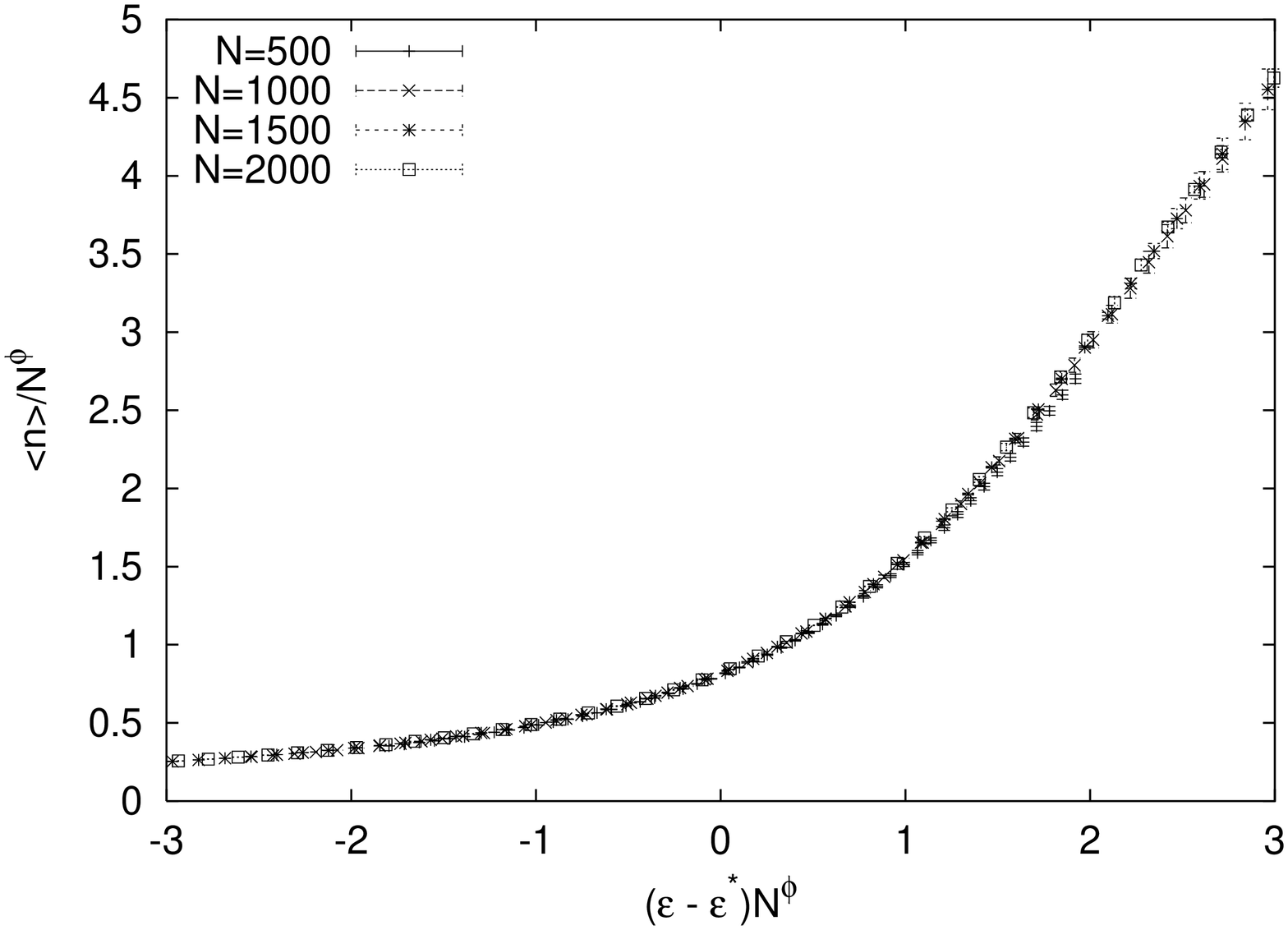,angle=270,width=7cm}
\epsfig{file=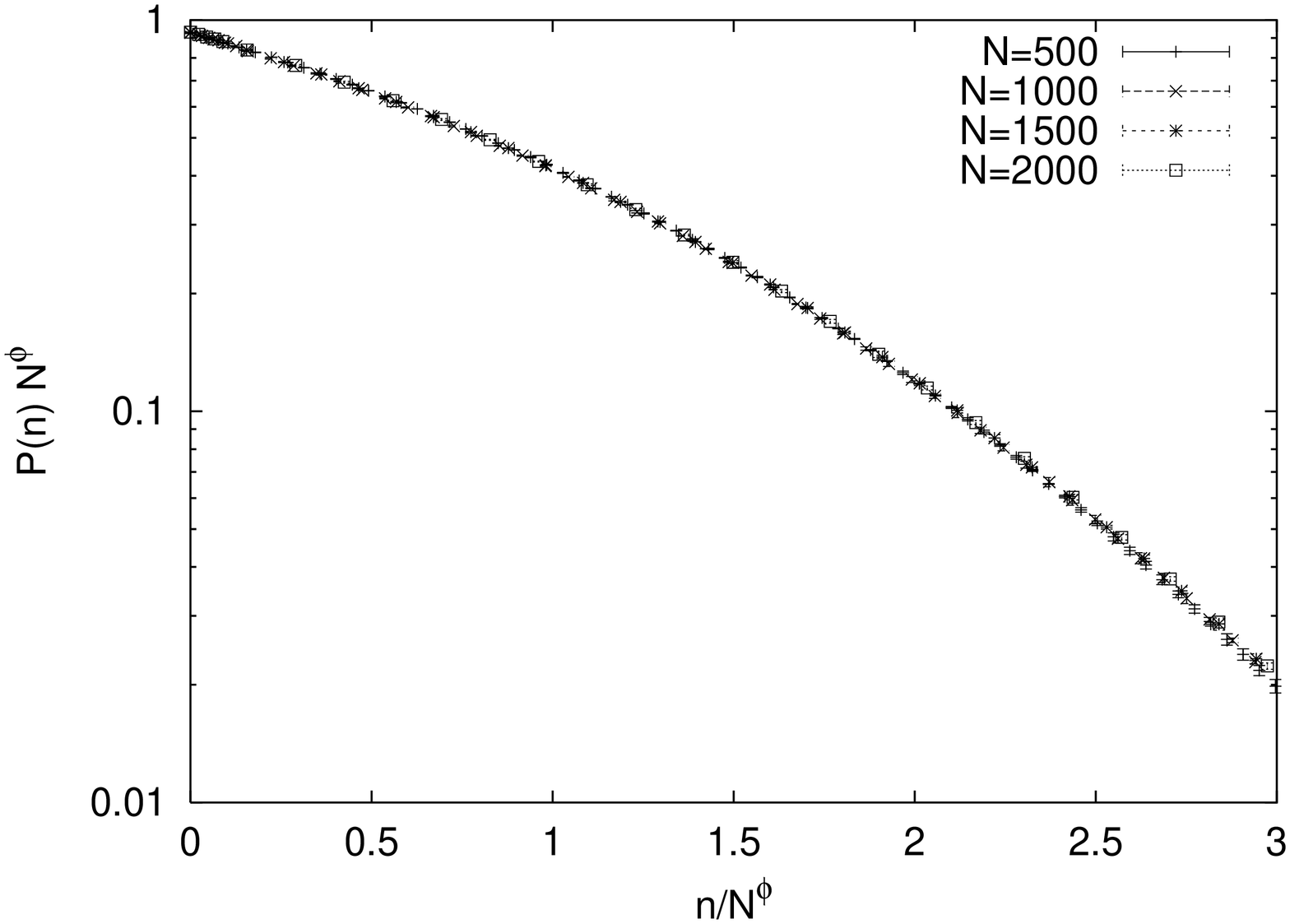,angle=270,width=7cm}}
\caption{Scaling plots of the average contact number for random walks in $3d$,
  and of the corresponding probability distribution at $\eps=\eps^*$. Here
  $\eps^*=1.07726$ and $\phi=1/2$ are the exact values (see
 Appendix). }
\label{eps_rw3d}
\vglue -14mm
\end{figure}

\subsubsection{d=3}

In $3d$ one finds a smooth second order transition when moving along the 
solid line in Fig.~\ref{phase-fig}, characterized by
a value $\phi=1/2$ of the crossover exponent. In Fig.~\ref{eps_rw3d}a we plot 
$\la n\ra /N^{\phi}$ as a function of $(\eps-\eps^*)N^{\phi}$.
The behaviour agrees with the expected scaling law, i.e. the data are
compatible with the exact values of $\eps^*$ and of $\phi$. Furthermore, 
finite size corrections appear to be small. The probability distribution
$P(n)$ exactly at the critical point, properly rescaled by $N^\phi$ and 
plotted against $n/N^{\phi}$, is shown in Fig.~\ref{eps_rw3d}b. The 
perfect data collapse confirms both the validity of the scaling law Eq.(\ref{PE})
and the efficiency of the numerical method.
 
\subsubsection{d=5}

\begin{figure}[htb]
\vglue -2mm
\centerline{\epsfig{file=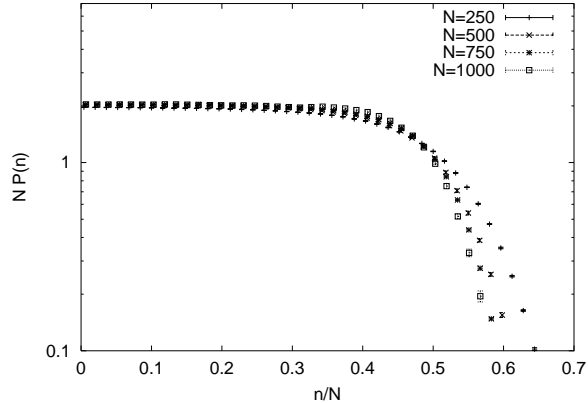,angle=270,width=8cm}}
\caption{Probability distribution of contact numbers 
for random walks in $5d$ at the exact critical value $\eps^*=2.00115$.}
\label{pe_tm_rw5d}
\vglue -3mm
\end{figure}

Let us now turn on the more intriguing $5d$ case. It is shown in the 
Appendix that the system undergoes a first order transition, since 
the contacts density $\langle n\rangle/N$ is discontinuous at $\eps^*$
in the thermodynamic limit, but shows scaling with a value $\phi=1$ 
of the crossover exponent.
We present in Fig.~\ref{pe_tm_rw5d} data for $P(n)$ at the analytically 
calculated critical value $\eps^*$. This plot is completely analogous to 
Fig.~\ref{eps_rw3d}b, but uses the exact value $\phi=1$ (see appendix). We 
see now definitely larger corrections to scaling. Also, the curves are 
slightly cap-convex which shows that $\langle n\rangle/N$ is not strictly 
discontinuous at $\eps=\eps^*$. But this is obviously a finite-size 
effect. For $N=1000$, the maximum of $P(n,\eps)$ jumps from $n/N=0$ to 
$n/N=c^*\approx 0.4$ when $\eps$ is increased from $\eps^*$ to $1.0005\eps$.
In the limit $N\to\infty$, on the basis of our numerical results
we can conjecture that $P(n,\eps^*)$ is flat
between $n/N=0$ and $n/N=c^*$.

On the other hand, it is obvious that $P(n)$ does not have the double-peak 
structure familiar from usual first order transitions. In these usual cases, 
the valley between the peaks is due to surface tension: States with the 
order parameter between the two peak values contain more than one domain 
and are therefore suppressed by surface tension. In the present case, there 
is no penalty for a transition between a molten and a double-stranded 
domain, explaining why all configurations with $0\leq n/N\leq c^*$ can be
equally much populated (for magnetic polymers, a similar scenario was 
proposed recently in \cite{vari}). Notice that
scaling as formulated in the previous 
section would not hold for usual first order transitions, since the valley 
between the peaks becomes exponentially deep in the thermodynamic limit, 
$P \sim \exp(-\sigma L^{d-1})$ for systems of linear size $L$. Formally, 
this can be reconciled with the present case by noticing that our 
polymers have topological dimension equal to one.

Related to this are strong fluctuations in the separation of the two chains 
in the double-stranded phase. Indeed, as proven in the appendix, the average the end-to-end 
diverges as $\xi_1 \sim (\eps-\eps^*)^{-1/2}$ as the critical point is 
approached from $\eps>\eps^*$. This shows that the `thermal' correlation length
exponent is $\nu_T=1/2=1/D$ and $\alpha_T=2-D\nu_T=2-\phi^{-1}=1$.

\begin{figure}[htb]
\vglue -5mm
\centerline{\epsfig{file=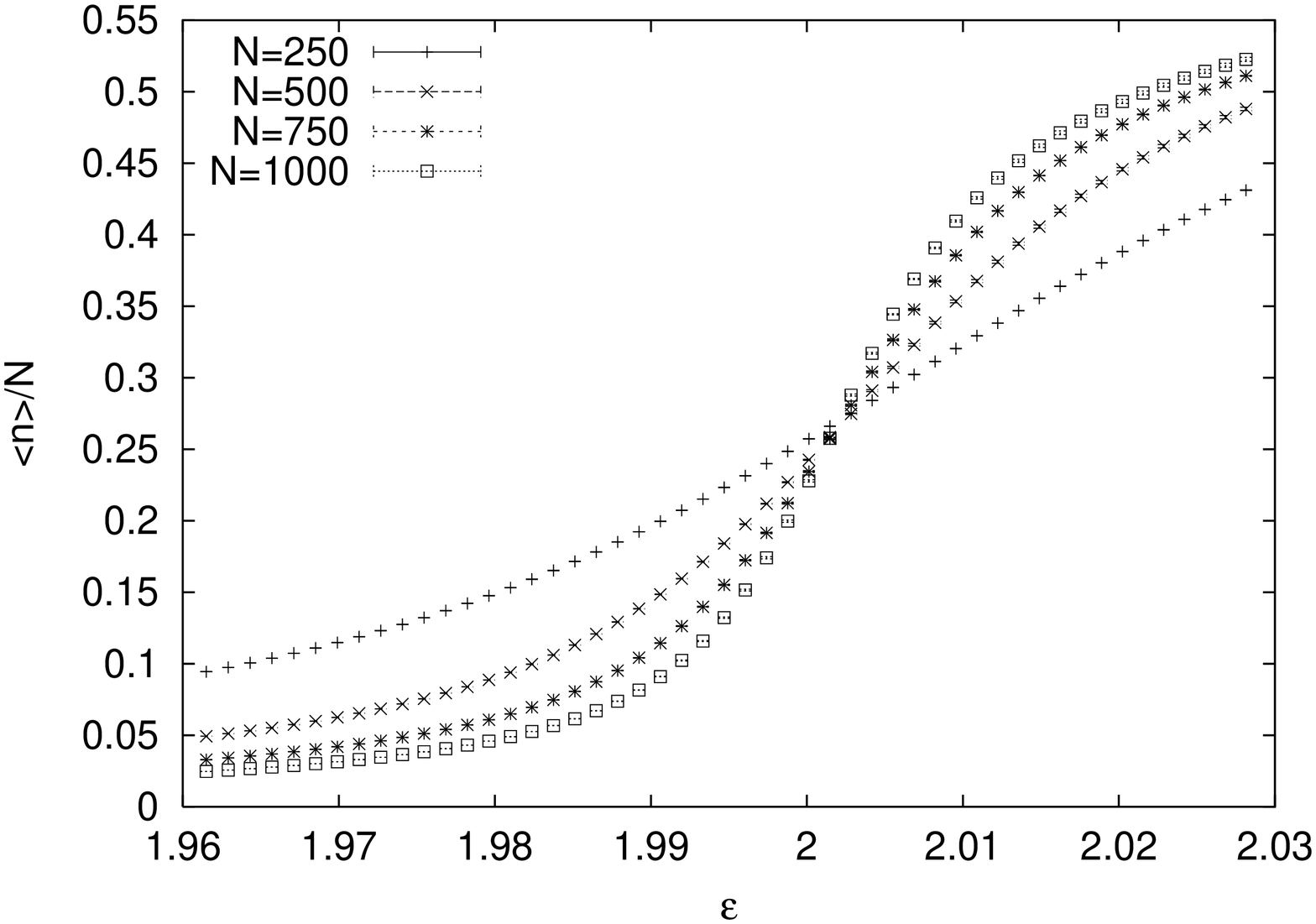,angle=270,width=6cm}
\epsfig{file=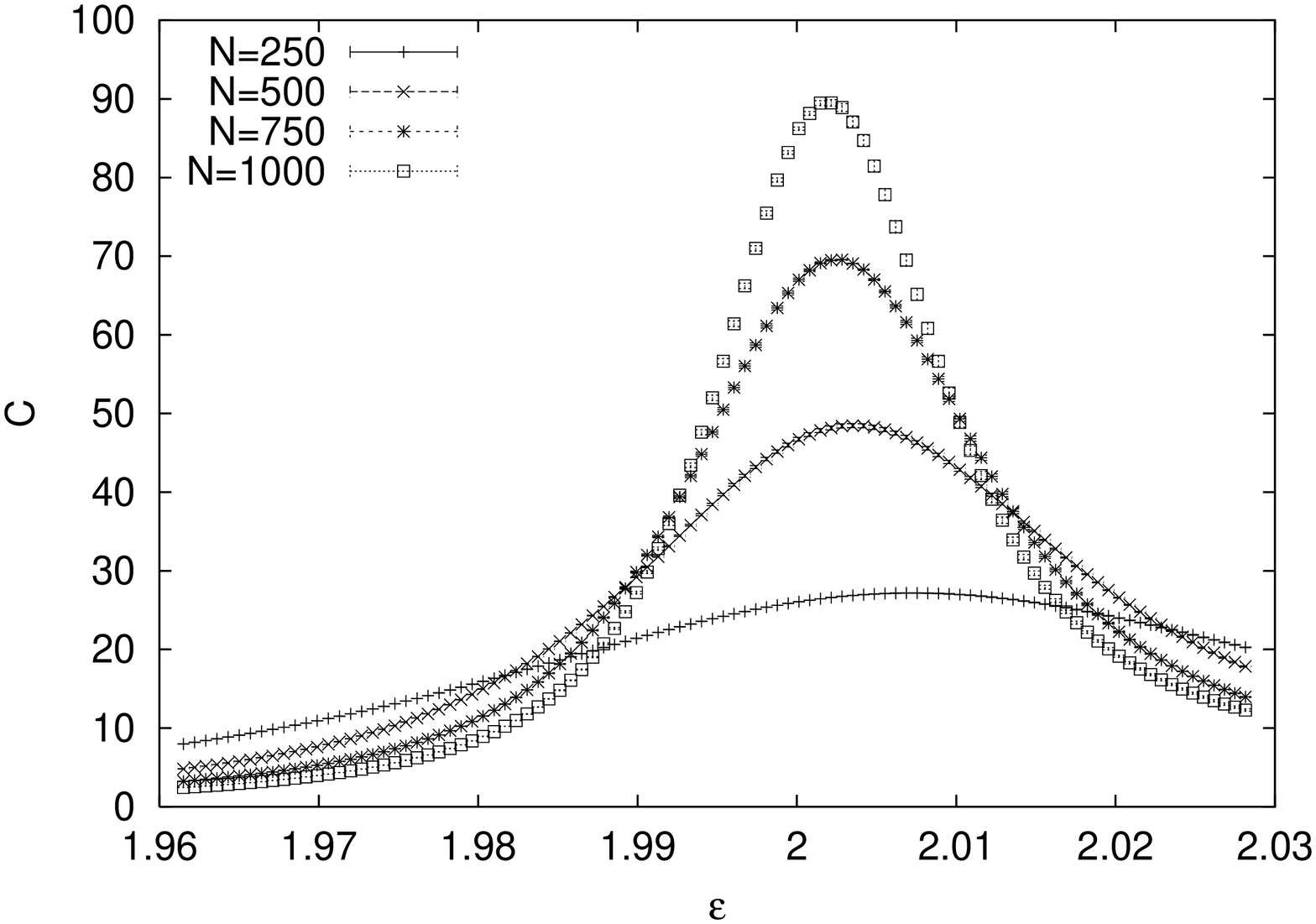,angle=270,width=6cm}}
\caption{Average contact number and specific heat for random walks in $5d$.}
\label{ec_rw5d}
\vglue -3mm
\end{figure}

In Fig.\ref{ec_rw5d} we plot the contact number density (left panel) 
and the specific heat (right panel), 
\be
   C = N^{-1}\frac{\partial E}{\partial T}=N^{-1}\eps^2 \frac{\partial 
      \langle n\rangle}{\partial \eps} = 
       N^{-1} \eps^2 [ \langle n^2\rangle - \langle n\rangle^2] ,
\ee
against $\eps$. In Fig.\ref{ec_rw5d}a, curves for different chain 
lengths considered cross nearly at the same $\eps$ value which is 
also the value the specific heat maximum is moving towards. Within 
the expected uncertainty, it agrees with the predicted $\eps^*$. But when 
trying to make the data collapse by plotting them against $(\eps-\eps^*)
N^\phi$, we see again strong corrections to scaling. 
Indeed, such fits 
using $\la n\ra/N^\phi$ would give $\phi\simeq 0.93$ instead of $\phi=1$. 
Another estimate for the crossover exponent can be recovered from the
way in which the maximum of the specific heat moves toward $\eps^*$
as the length of the walk increases.
Since in the crossover region the specific heat follows the crossover scaling
the position of the maximum for $N$ finite is shifted by a term proportional
to $N^{-\phi}$.
$\phi >1$ by a small amount. These deviations from perfect 
scaling give us a hint of what we have to expect when going now over to 
self avoiding walks.

\subsection{Self Avoiding Walks, $d=3$}

\subsubsection{Scaling of $P(n)$ and Properties Derived from it}

\begin{figure}[htb]
\vglue -1mm
\centerline{\epsfig{file=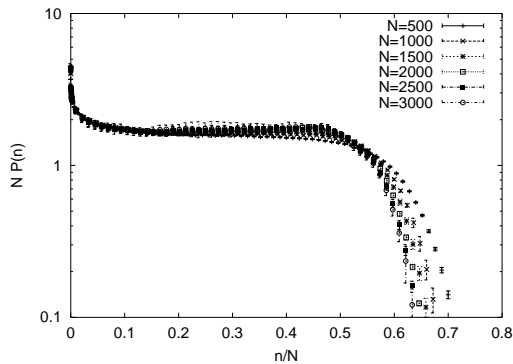,angle=270,width=7cm}}
\caption{Probability distribution of the contact numbers
   for SAWs in $3d$ at the estimated critical value $\eps^*=1.3413(4)$. }
\label{pe_tm_saw3d}
\vglue -6mm
\end{figure}

\begin{figure}[htb]
\vglue -1mm
\centerline{\epsfig{file=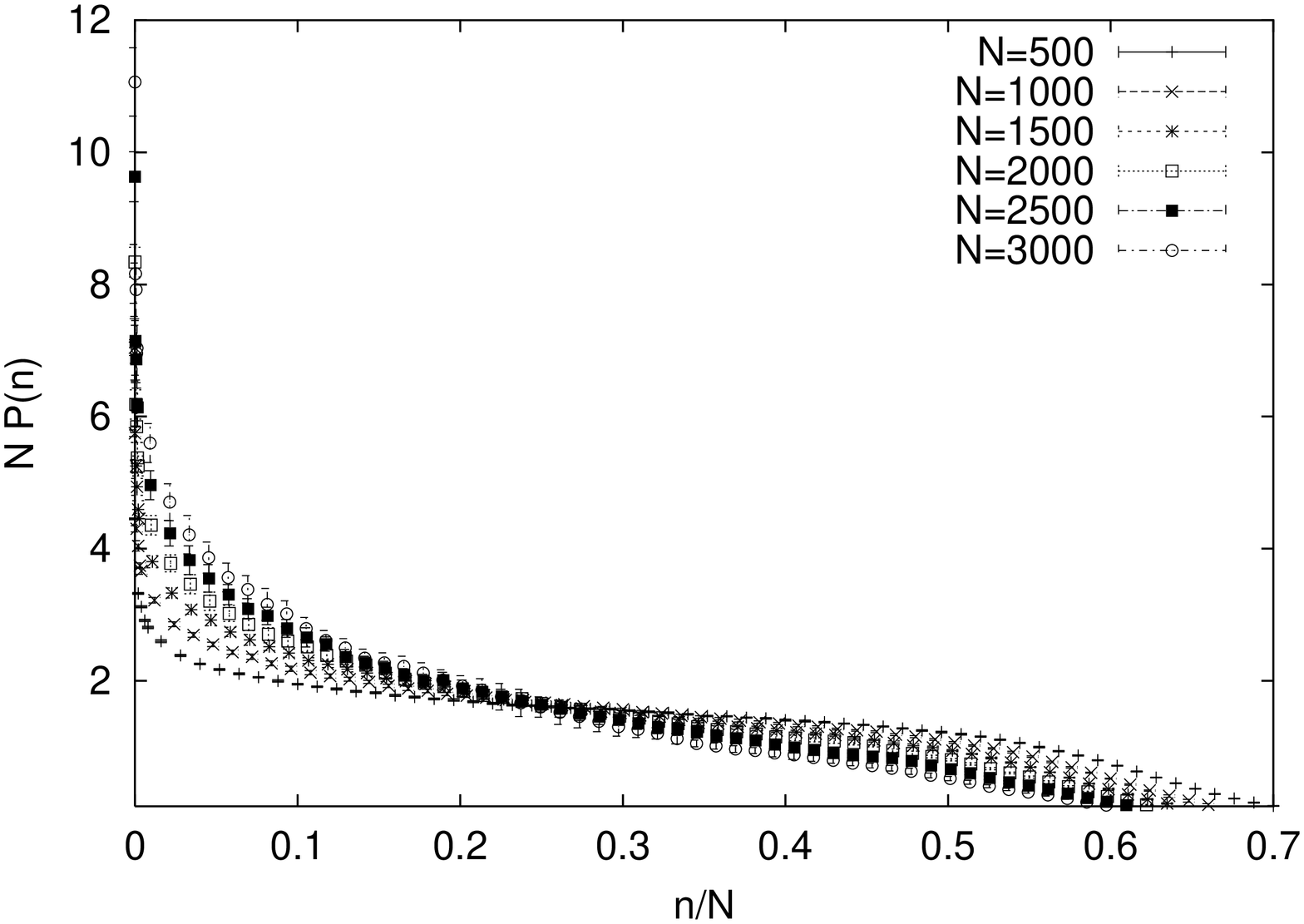,angle=270,width=7cm}
\epsfig{file=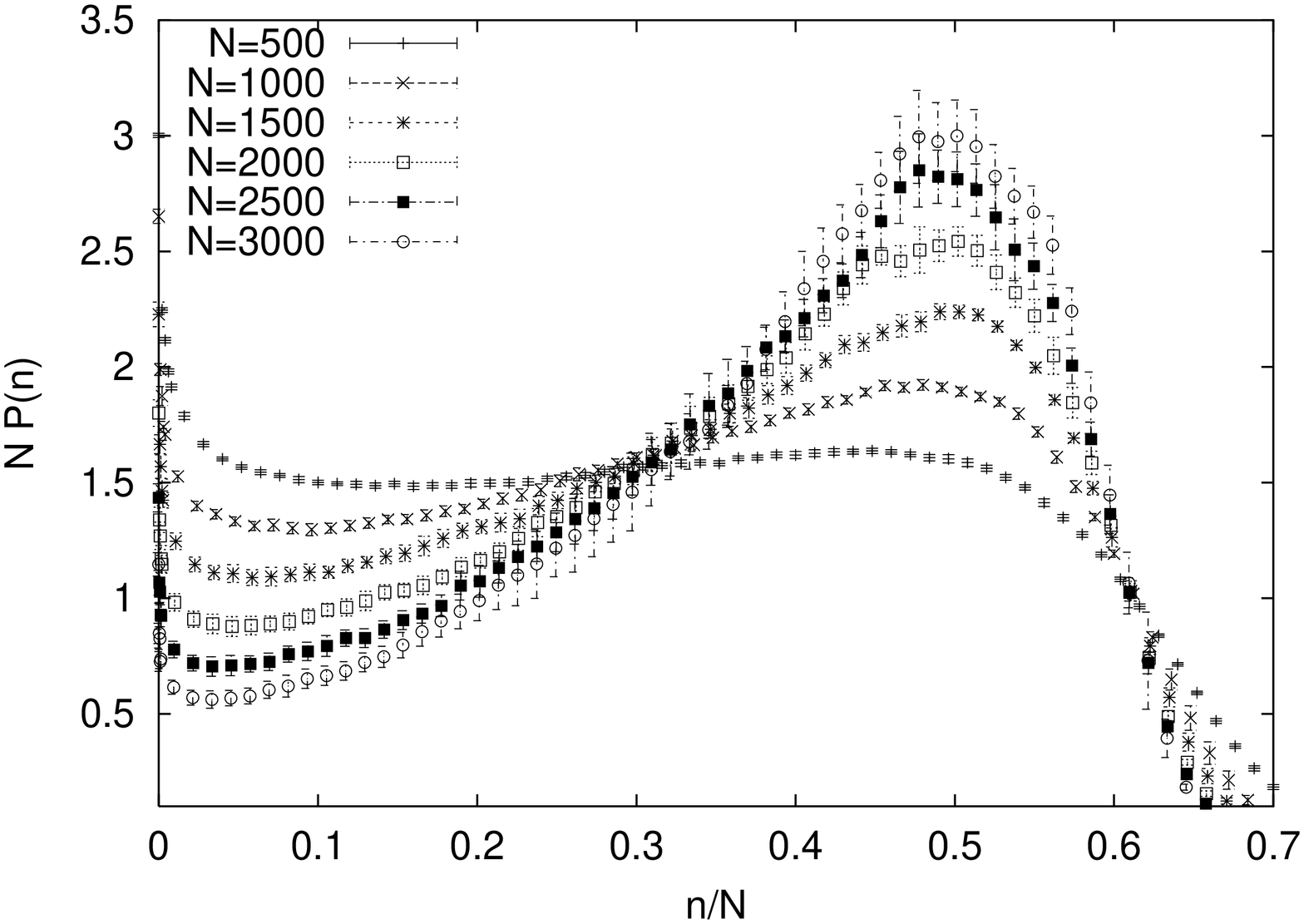,angle=270,width=7cm}}
\caption{Probability distribution of the contact numbers 
  for SAW's in $3d$ at $\eps=.999 \eps^*$ and $\eps=1.001 \eps^*$. }
\label{p_pm_tm_saw3d}
\vglue -4mm
\end{figure}

In Fig.~\ref{pe_tm_saw3d} we show $\log P(n,\eps^*)$, where 
$\eps^*=1.3413\pm 0.0004$ is our best
estimate of the critical $\eps$ value. This distribution is
clearly not convex, in contrast to the case of 5-dimensional ideal 
random walks studied in the previous subsection. Instead, there is 
a peak at $n=0$. Due to it, the maximum of $P(n,\eps)$ jumps 
discontinuously when $\eps$ passes through $\eps^*$ (see 
Fig.~\ref{p_pm_tm_saw3d}). Apart from this, the situation is 
very similar. We see again substantial corrections to scaling, 
but it seems quite clear that scaling works with $\phi=1$. In particular, 
the depth of the valley between the peak at $n=0$ and the shoulder at 
$n/N=c^*\approx 0.5$ does not increase with $N$. And, what is more 
important, the value of $c^*$ does not substantially decrease with 
$N$. This is our strongest numerical evidence for the transition to 
be of first order. It should be noted that here we are
considering up to quite long chains ($N=3000$) and that the data for
$N=2500$ and for $N=3000$ are nearly indistinguishable.

Deviations from the scaling behavior are seen mostly for large $n/N$.
There, the distribution becomes increasingly steeper with $N$. This 
was seen also for ideal random walks in $d=5$. It 
is indeed easy to understand. At large $n$, we expect $P(n) \sim 
e^{\eps n} \li N/n \ri^n \li 1- n/N\ri^{n-N}$ which does 
not follow our scaling law but is in qualitative agreement with our 
data.

\begin{figure}[htb]
\vglue -.1cm
\centerline{\epsfig{file=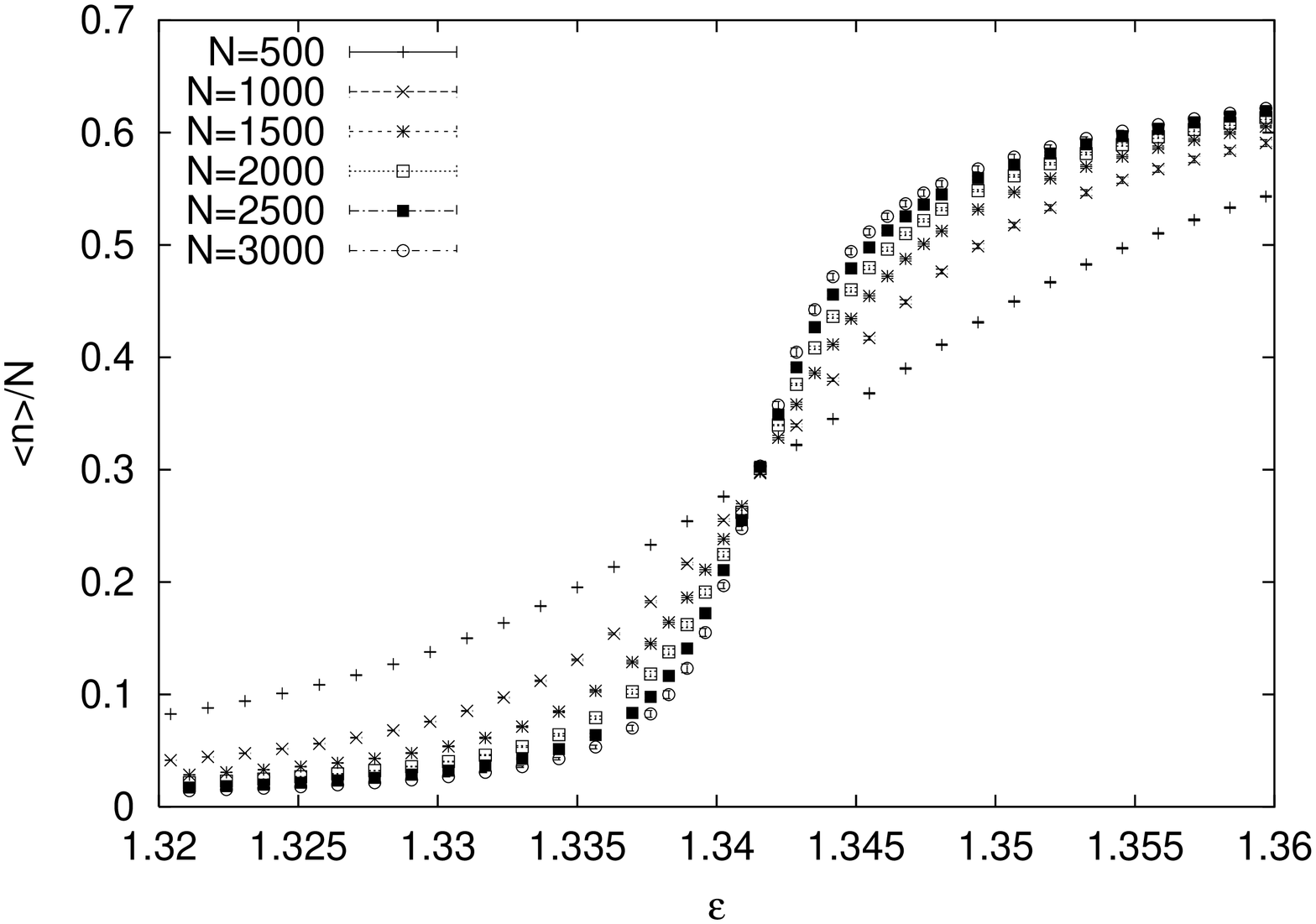,angle=270,width=7cm}
\epsfig{file=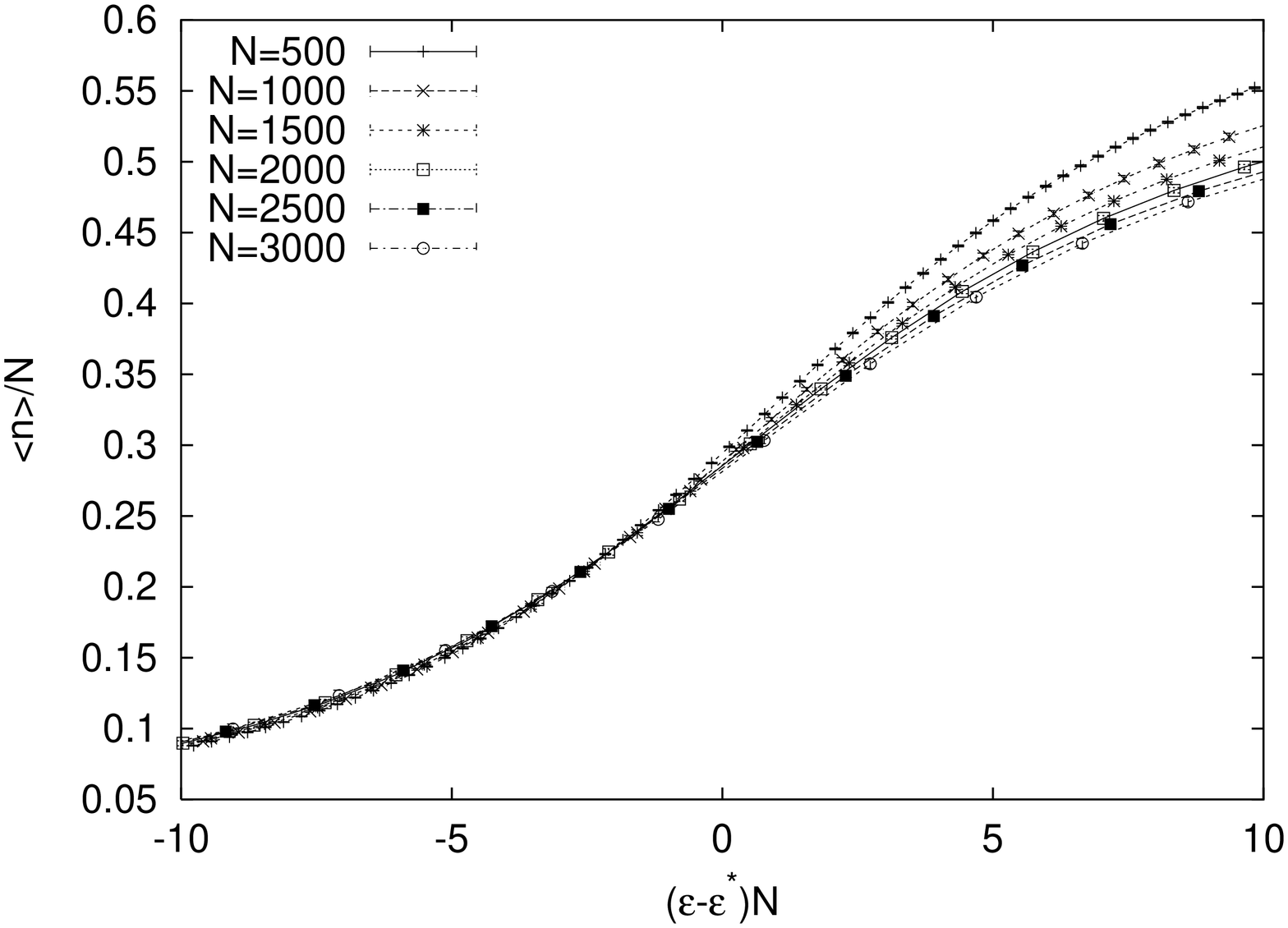,angle=270,width=7cm}}
\caption{The average contact number plotted against $\eps$ (left panel) and against 
   $(\eps-\eps^*)N$ (right panel).  }
\label{ec_saw3d}
\vglue -6mm
\end{figure}

Accordingly, also the scaling of $\la n\ra/N$ becomes poor for large 
$\eps$, as seen from Fig.~\ref{ec_saw3d}b. If we would try to optimize 
this scaling plot, we would find $\phi\approx 0.94$. But we know that 
this would be wrong since this would put too much emphasis on the region 
$n/N\approx 1$ in the contact number distribution which we know to be 
not scaling.

\begin{figure}[htb]
\vglue -.1cm
\centerline{\epsfig{file=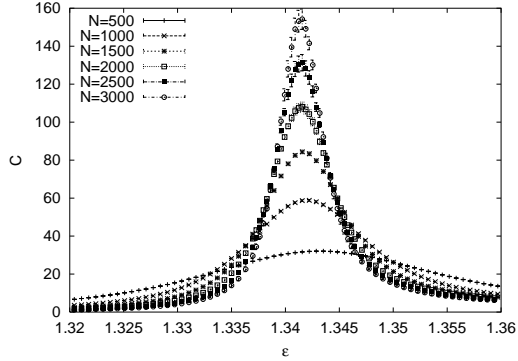,angle=270,width=7cm}}
\caption{The specific heat.}
\label{ec_saw3d-C}
\vglue -3mm
\end{figure}

\begin{figure}[htb]
\vglue -1mm
\centerline{\epsfig{file=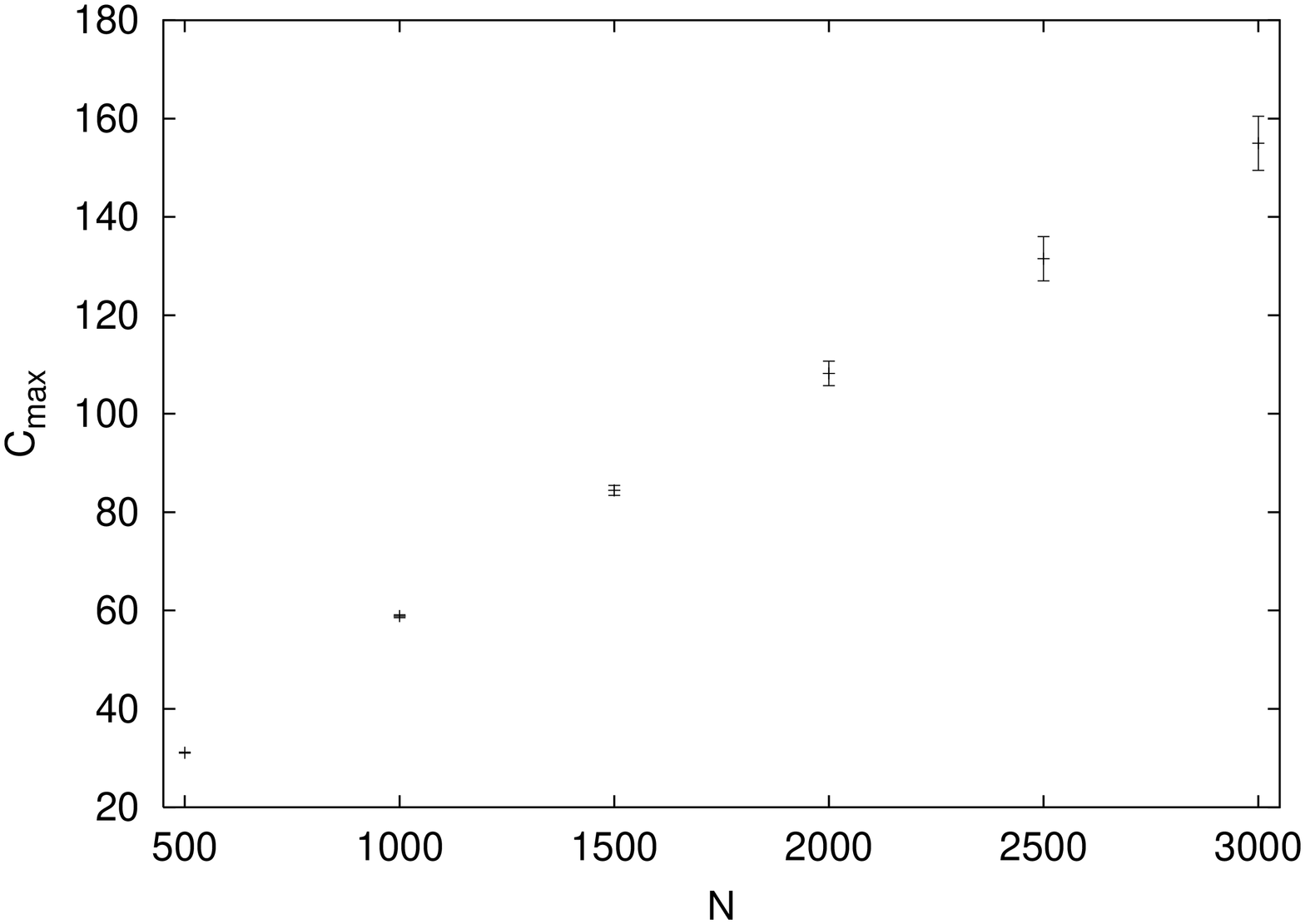,angle=270,width=7cm}}
\centerline{\epsfig{file=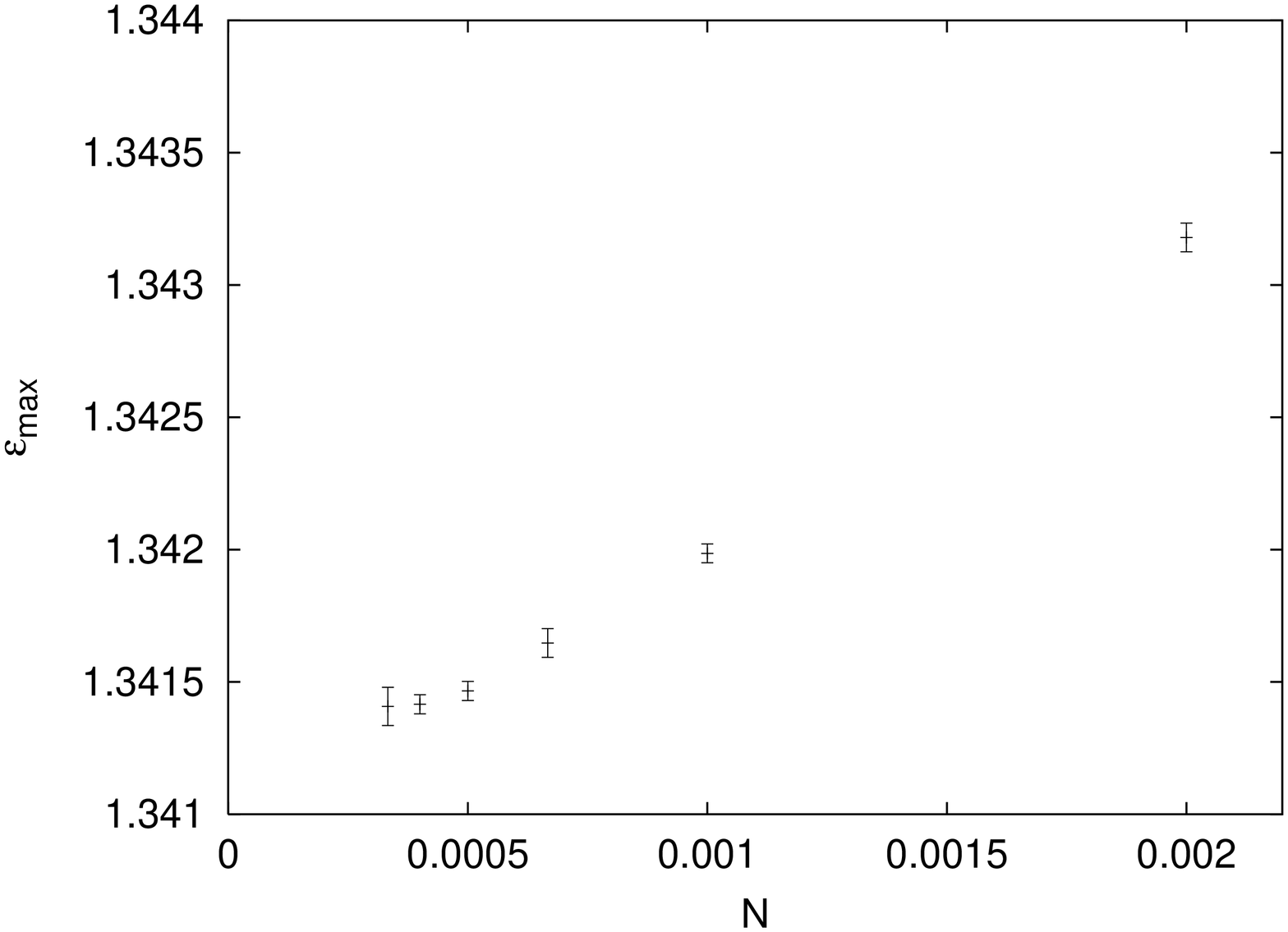,angle=270,width=7cm}}
\caption{The specific heat maximum and the corresponding $\eps$
  value for SAW's in $3d$.}
\label{ec_rw5d_zzz}
\vglue -9mm
\end{figure}

We face a similar situation when looking at the specific heat. As seen 
in Fig.~\ref{ec_saw3d-C}, the height of the  maximum (see Fig.~\ref{ec_rw5d_zzz}a) increases roughly 
$\propto N$, which would correspond to $\phi=1$, but a least squares fit of our data
shows an effective exponent $\phi < 1 $.
Since we are working with $N\le 3000$, in a regime which is far from the critical one, 
we expect that our estimate is biased by the presence of corrections to scaling.
Anyway no clear trend in the computed exponent is
obtained from fits of data with cuts corresponding to increasing values of $N$
(see table \reff{tabella}). All the obtained values are compatible in the statistical errors,
but a $\chi^2$ analysis indicates that our errors could be overestimated.
Nevertheless the central value of the effective exponent increases with the increasing
of $N_{min}$. This could indicate the presence of corrections to scaling. 
The last value obtained for $N=1500$ is compatible with $\phi=1$.

\begin{table}
\protect\footnotesize
\begin{center}
\begin{tabular}{|c|c|c|c|}
\hline
\hline
$N_{\rm min}$ & $ \phi_{\rm eff} \pm \Delta\phi_{\rm eff} $& $\chi^2$ & DF\\
\hline
  500 &     $     0.920 \pm 0.019  $&    0.042  &  3\\
  1000&     $     0.925 \pm 0.055  $&    0.039  &  2\\
  1500&	    $	  0.98  \pm 0.15   $&    0.0015 &  1\\
\hline
\end{tabular}
\caption{Effective crossover exponent $\phi_{\rm eff}$ from the fits of the maximum of the
specific heat as a function of minimum value of $N_{\rm min}$ considered in the fit.} 
\label{tabella}
\end{center}
\end{table}

On the other hand, fitting the shift of the position of the
maximum, which is shown in Fig.~\ref{ec_rw5d_zzz}b, one would get $\phi=1.31(14)$.
In this case the large statistical errors on our data do not allow any guess on the
correction to scaling effects.


In summary, these results show that the melting transition for 
interacting SAWs in 3 dimensions is first order, while the analogous transition 
for interacting ideal walks in 3 dimensions is second order. This agrees 
with our expectation that excluded volume effects should make the transition 
sharper. We should add that we also performed simulations of a third version 
of the model, in which the two polymers were self- but not mutually avoiding 
(data not shown). 
While two monomers of the same chain were not allowed to occupy the same site, 
two monomers from different chains were allowed to do so, and this contributed 
to the energy if and only if the indices of the monomers were the same. For this 
intermediate model we found a $P(n)$ very similar to Fig.~\ref{pe_tm_saw3d}, 
but with a less pronounced peak at $n=0$ and a very small non-convex region.
Therefore we are not sure whether the transition in this model is first or 
second order.

\subsubsection{The Thermal Correlation Length}

Up to now we only looked at the scaling behavior of $P(n)$ and at quantities 
which are related to it straightforwardly. We mentioned already in Sec.~\ref{scl} 
several scaling laws which are less directly related.

\begin{figure}[htb]
\vglue -2mm
\centerline{\epsfig{file=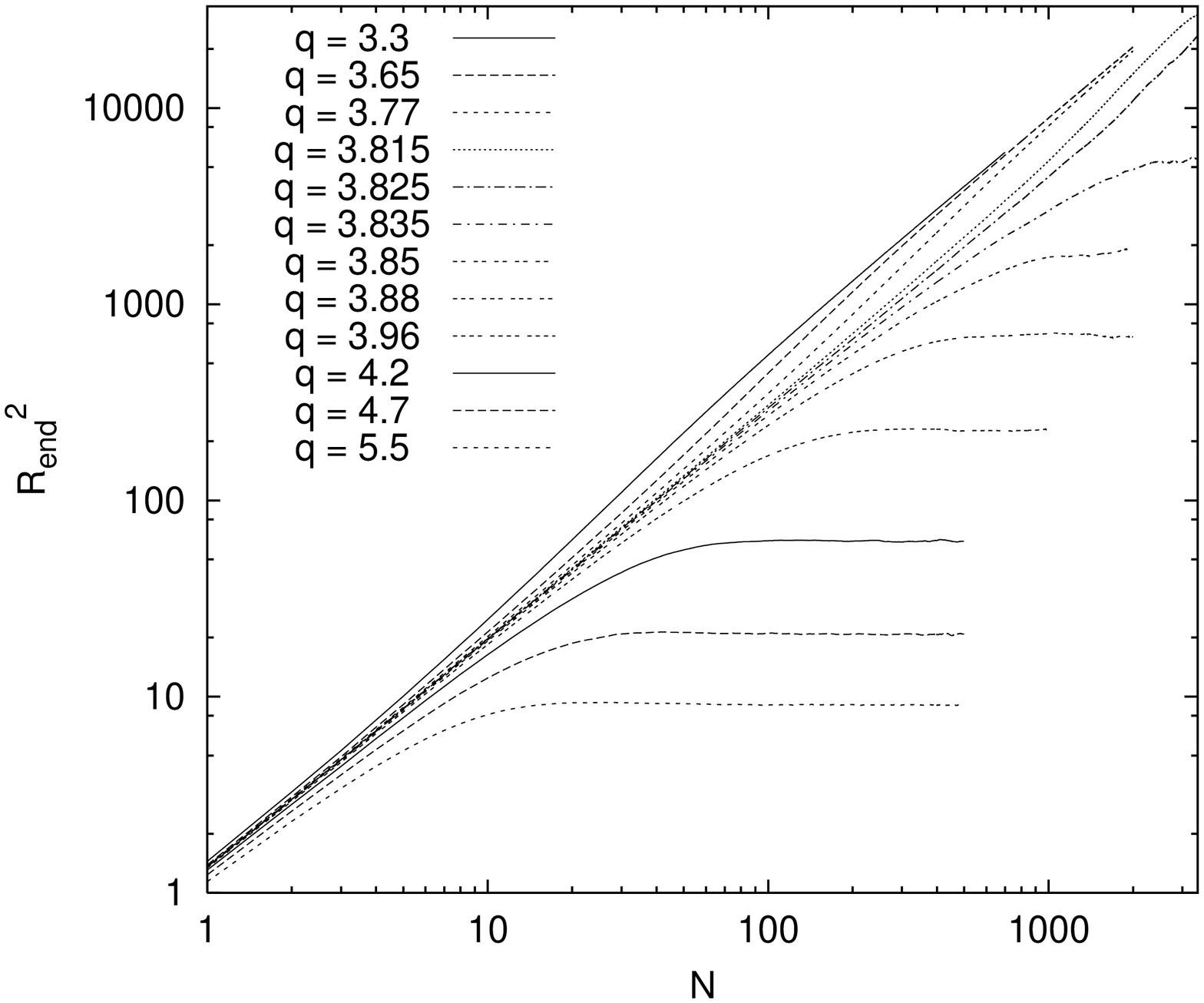,angle=270,width=8cm}}
\caption{Average squared end-to-end distance $R_{\rm end}^2$ for various values of 
  $q=e^\eps$, plotted against $N$ on a double logarithmic scale. Since all curves 
  are based on independent runs, their typical fluctuations relative to each other 
  indicate the order of magnitude of their statistical errors.}
\label{fig:R-end}
\vglue -9mm
\end{figure}

\begin{figure}[htb]
\centerline{\epsfig{file=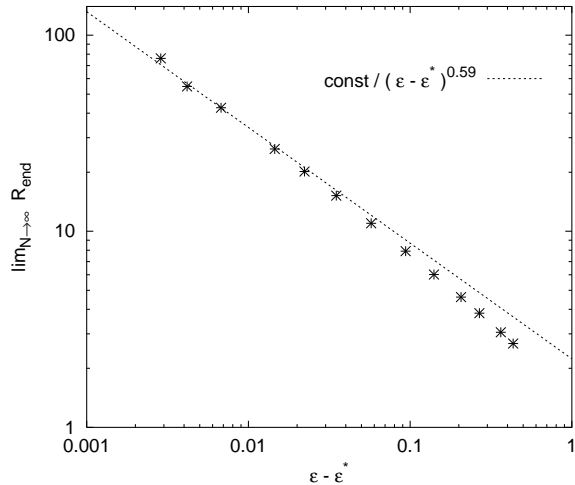,angle=270,width=8cm}}
\caption{Thermal correlation length $\xi_1=\lim_{N\to\infty} R_{\rm end}$ plotted 
  against $\eps - \eps^*$ on a log-log plot. For large values of $\eps - \eps^*$,
  statistical errors are much smaller than the symbols. For $\eps - \eps^* < 0.01$
  we cannot give exact errors, but rough estimates can be obtained by 
  comparing with Fig.~\ref{fig:R-end}. The dashed line has slope -0.59, 
  corresponding to $\nu_T=\nu$. }
\label{fig:R-end-scal}
\vglue -12mm
\end{figure}

\begin{figure}[htb]
\centerline{\epsfig{file=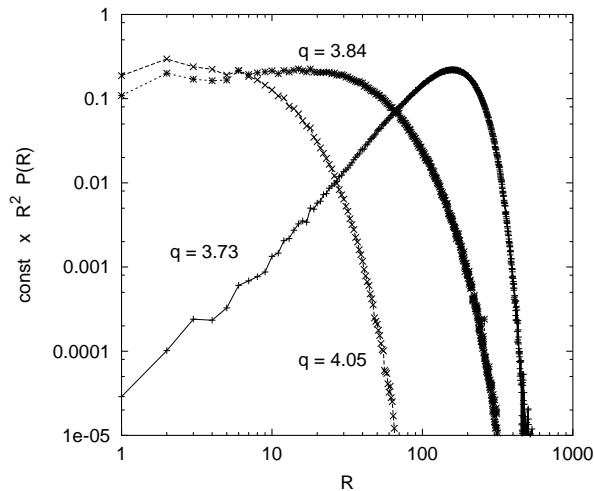,angle=270,width=8cm}}
\caption{Distributions of end-to-end distances at $N=3000$ for three values of $q=e^\eps$. 
   Normalization is arbitrary. }
\label{fig:P-R}
\vglue -5mm
\end{figure}

\begin{figure}[htb]
\centerline{\epsfig{file=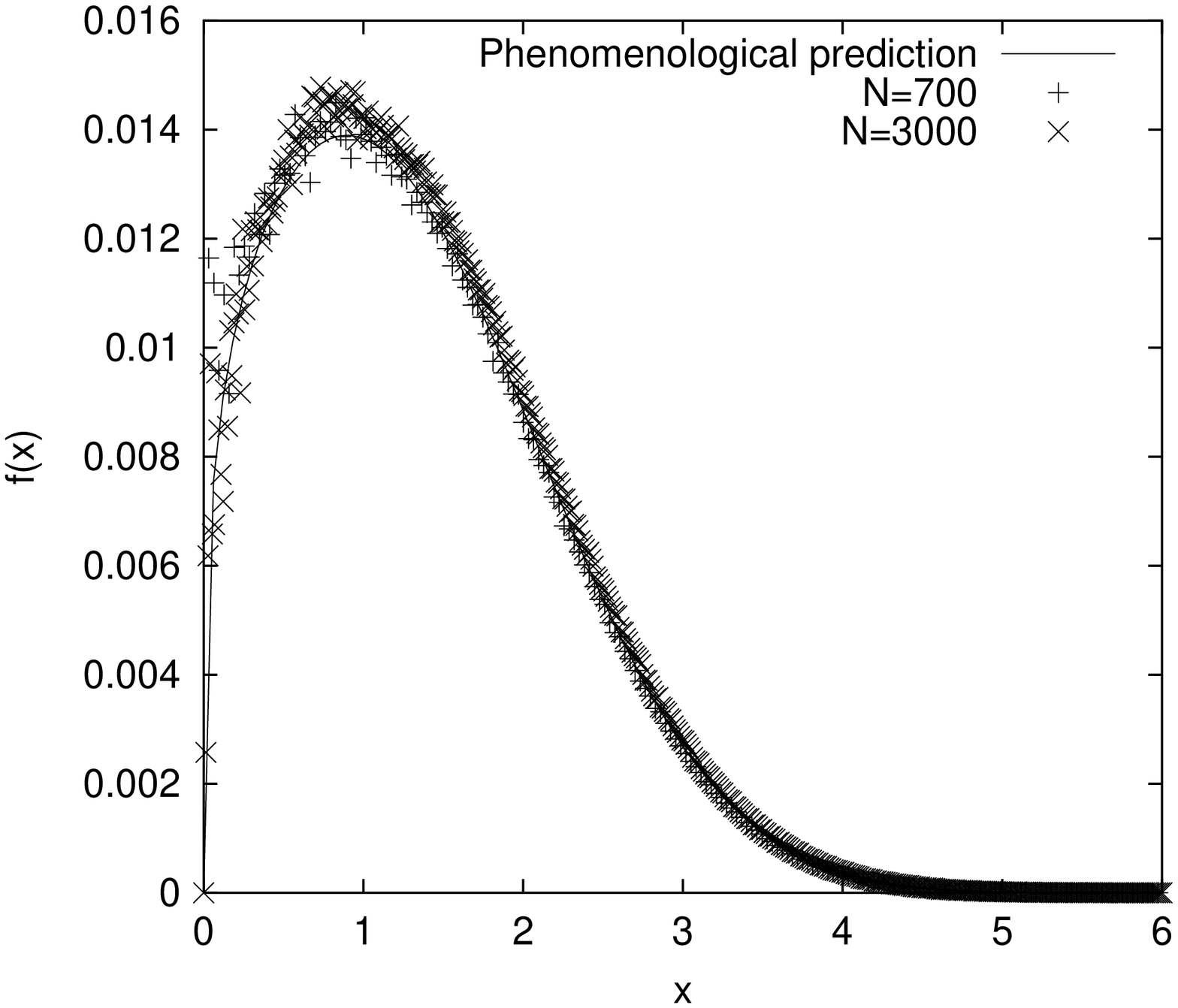,angle=270,width=8cm}}
\caption{Comparison of the scaling functions $f_N(x)$ for $N=700$, $\eps=1.19$ and $N=3000$,
$\eps=1.32$ with the phenomenological prediction for $f(x)$ of \cite{phen1,phen2}.}
\label{comp_dist}
\vglue -12mm
\end{figure}

The most interesting concern the {\em thermal} correlation length (see Sec.4). 
In the double-stranded phase ($\eps>\eps^*$) we can identify it with the 
rms. distance $R_{\rm end}$ between the endpoints of the two chains. Numerical
results for the latter (both for $\eps>\eps^*$ and for $\eps<\eps^*$) are 
shown in Fig.~\ref{fig:R-end}.  For $\eps>\eps^*$,
$R_{\rm end}$ tends for $N\to\infty$ to a constant which diverges when the
tricritical point is approached, showing that there is indeed a divergent
thermal correlation length which is independent of the system size for large
systems. Exactly at the tricritical point we find $R_{\rm end}\sim N^\nu$,
showing that there the thermal and the ``geometrical" correlation length
(the Flory radius) coincide. The latter is also known from polymer adsorption
to a wall, e.g. \cite{eisenriegler,hegger}. The divergence of $\xi_1 = 
\lim_{N\to\infty} R_{\rm end}^2$ for $\eps \to \eps^*+0$ is shown 
in Fig.~\ref{fig:R-end-scal}. We see that the {\em thermal} correlation length 
exponent $\nu_T$, defined by $\xi_1 \sim (\eps - \eps^*)^{-\nu_T}$, agrees 
with the Flory exponent $\nu$ (the {\em geometrical} correlation length exponent).
Since $\phi=\nu/\nu_T$ \cite{dosSantos,Derrida}, we find again $\phi=1$.

For $\eps<\eps^*$ we also have $R_{\rm end}\sim N^\nu$. But in this regime, 
$R_{\rm end}$ does not scale as the thermal correlation length. Instead, 
we can identify $\xi_1$ with the average diameter of small molten `bubbles'
(which we did not measure since our algorithm would give very large errors). 
Since SAWs in 3 dimensions are not recurrent, large bubbles do not occur 
in the molten phase, and $\xi_1$ is finite and decreases with decreasing
$\eps$.

Finally, end-to-end distance distributions are shown in Fig.~\ref{fig:P-R}.
For $\eps < \eps^*$  they coincide for large $N$ with the end-to-end 
distance distributions of non-interacting SAWs of length $2N$, except for a 
region of very small distances which becomes irrelevant in the limit 
$N\to\infty$. More precisely, if we denote with $c_N(\vec R)$ the 
number of $N$-step walks whose end-points 
are at a distance $R=|\vec R|$ apart, we have 
\be
   f_N(x) \equiv R_{\rm end}^d P_N( \vec R) = 
   R_{\rm end}^d \frac{c_N(\vec R)}{\sum_{\vec R} c_N(\vec R)}
   \approx f(x) \lc 1+O \li N^{-\Delta} \ri\rc
\ee
where $x = (2d)^{1/2} R /R_{\rm end}$, $f(x)$ is a universal 
function, and $\Delta$ is a correction to scaling exponent. As shown in 
\cite{distr}, $f(x)$ is well approximated by a phenomenological representation
given first by Mc Kenzie and Moore \cite{phen1,phen2}. A comparison 
with the latter is shown in Fig. \ref{comp_dist}. For small $x$ the attraction
between monomers is felt and $f_N(x)$ is larger than for ordinary SAWs,
but this effect disappears for 
$N\to\infty$, as long as $\eps$ is strictly smaller than $\eps^*$.
The transition between the regimes $\eps>\eps^*$ and $\eps<\eps^*$ is not 
through a double-peaked 
distribution as one might expect for a first order transition, but there is 
an approximately flat plateau at $\eps \approx \eps^*$.

\subsubsection{The molten / double stranded phase transition}

The boundary between the two dense phases in Fig.~1 (crosses) was obtained 
by taking finite lattices of size $L^3$ with periodic boundary conditions. 
During the simulation we measured only the energy (number $n$ of contacts) and 
the partition sum. The phase transition is seen as a rapid increase of $\la n\ra$
when the monomer density $\rho = N/L^3$ increases above a value $\rho_1(\eps)$.
Let us define $z_1(\eps)$ such that
\be
   {\partial \over \partial N }
      \left[ z_1(\eps)^N Z_N(\eps) \right]_{N=\rho_1(\eps)L^3} = 0.
                       \label{z_1}
\ee
If the transition were second order, $z_1(\eps)$ would be the critical 
fugacity at the considered value of $\eps$. But our simulations show 
clearly that the transition is first order. In this case, there is a second 
value of the density $\rho_2(\eps)> \rho_1(\eps)$ such that the system is 
in the double-stranded phase when $\rho>\rho_2$, in the molten phase when 
$\rho<\rho_2$, and in a mixed phase in between. Equation~\reff{z_1} holds 
in the entire interval $\rho_1<\rho<\rho_2$.

For most values of $\eps$ we used $L=32$, only for $\eps>1.3$ ($q\geq 3.7$) we 
used larger lattices of size up to $64^3$. Using the algorithm as described 
in Sec.4 we were able to see the first threshold $\rho_1$, but not the 
second one. Obviously, our algorithm is not efficient enough.
Also, the increase of $\la n\ra$ was too slow for $\rho>\rho_1$. On the 
other hand, lattices of different sizes gave roughly consistent values of 
$\rho_1$. The algorithm became much more efficient when we based the ``population
control" (copying and pruning) on weight factors calculated with a larger 
value of $\eps$ than that used for evaluating average values (we used 
mostly $\eps=\eps^*$, but slightly larger values also worked well). This is 
easily understood. When the density is high enough that double-stranded 
configurations would be favored over molten ones, our ensemble just has 
no such configurations, with overwhelming probability. Thus the parts 
of the chains grown at ambient densities $\rho>\rho_1$ are correct, but 
the older parts are wrong and have virtually no chance to be corrected by 
re-growing. This is no longer so when we base the population control on a 
higher value of $\eps$. Then chains with larger $n$ are favored already 
from the very beginning. As long as $\rho<\rho_1$, they will not contribute 
significantly because they have a tiny weight. But at $\rho>\rho_1$, 
their typical weight will become larger than that of the completely molten chains, 
and they will start to contribute. We found that $\la n\ra$ indeed increased 
very rapidly in a very narrow density interval of a few percent, when using 
this improved algorithm. This still did not allow us to measure reliably 
the difference $\rho_2-\rho_1$, but it convinced us of the correctness of 
the scenario (in particular of the first order of the transition), and it 
gave uncertainties roughly as large as the symbol sizes in Fig.~1.

\subsubsection{The short chains / double stranded phase transition}

\begin{figure}[htb]
\vglue -3mm
\centerline{\epsfig{file=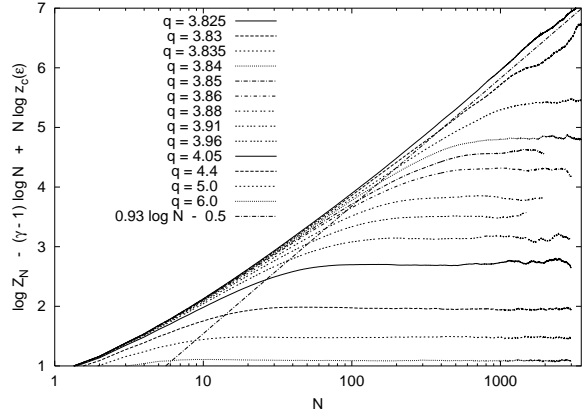,angle=270,width=8cm}}
\caption{Logarithms of partition sums at different values of $q=e^{\eps}$, 
after subtracting suitable terms so that 
they become flat for $N\to\infty$ and $\eps > \eps^*$. The straight dashed line 
has slope $\gamma^* - \gamma = 0.93$.}
\label{fig:Z}
\vglue -2mm
\end{figure}

\begin{figure}[htb]
\centerline{\epsfig{file=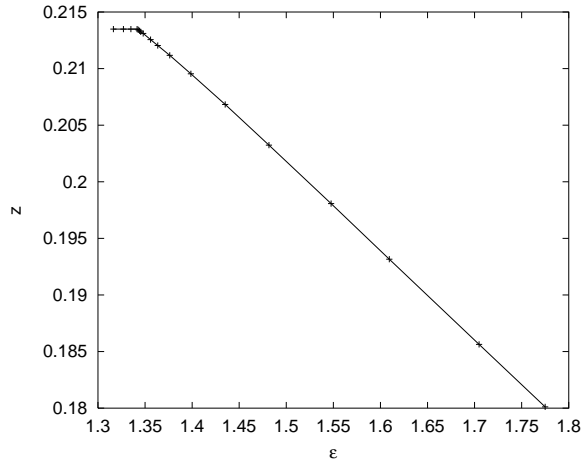,angle=270,width=8cm}}
\caption{Phase boundary between the short chain and double-stranded 
  phases.}
\label{phase-bdry-blowup}
\vglue -11mm
\end{figure}

In contrast to the above, the determination of the phase boundary between 
the phase having only short chains and the dense phase for $\eps>\eps^*$ 
was straightforward. Values $z_c(\eps)$ were obtained by plotting $\log Z_N(\eps) 
-(\gamma-1)\log N + N\log z$ against $\log N$ and changing $z$ until 
this became horizontal for large $N$. Several such curves are drawn in 
Fig.~\ref{fig:Z}. Notice that we cannot expect them 
to be flat at small values of $N$, since the behavior $Z_N \sim \mu^N N^{\gamma-1}$
is expected only when the length of the (double-stranded) chain is much larger 
than the typical size of a molten `bubble' and this makes difficult the determination
of the asymptotic regime. But this does not affect the uncertainty of $z_c(\eps)=1/\mu$ 
in a dramatic way, since $Z_N$ is extremely sensitive to even tiny changes of $\mu$. As 
a result we cannot give formal error estimates, but they are definitely 
smaller than $\pm 0.00005$. This number refers to $\eps\approx \eps^*$ 
where errors are largest, for $\eps\gg \eps^*$ our estimates are indeed much 
more precise. A blow up of this phase boundary is shown in 
Fig.~\ref{phase-bdry-blowup}. From that we verify that the boundary terminates 
at $(\eps^*,z^*)$ with finite non-zero slope, showing again that $\phi=1$.
For very large values of $\eps$ we see also that $[\mu(\eps)]^2 e^\eps \to \mu(0)$ 
as we should expect for a very tightly bound double-stranded chain.

{From} the curves with $\eps\approx \eps^*$ in Fig.~\ref{fig:Z} we can read 
off the exponent $\gamma^*$ controlling the partition sum exactly at the 
tricritical point. In view of the substantial finite size corrections at 
small $N$ and the uncertainties at large $N$ due to statistical errors and 
the uncertainty in the exact value of $\eps^*$, we obtain a rather crude 
estimate
\be
   \gamma^* = 2.09 \pm 0.1\;.
\ee
Unfortunately, we do not have any prediction for $\gamma^*$ to compare this 
with. But for ideal random walks we have $\sigma_{rw} = -1$ (see Appendix), 
where $\sigma$ was defined in Sec.4 and is related to $\gamma^*$ by Eq.\reff{sigma}.
Trying $\sigma=\sigma_{rw}$ as a first guess, we would predict $\gamma^*=2.16$ 
in surprisingly good agreement with our direct estimate.

Finally, we can read off from Fig.~\ref{fig:Z} values of the function
$A(\eps)$ in the scaling ansatz
\be
   Z_N(\eps) \simeq A(\eps) \mu(\eps)^N N^{\gamma-1} \;.
\ee
According to Eq.\reff{Z-scaling}, this function 
should scale as $A(\eps) \sim (\eps-\eps^*)^{\gamma-\gamma^*}$.
This was reasonably well satisfied. 

\section{Conclusion}

We studied a simplified model for DNA denaturation. It can be considered
as a lattice realization of the Poland-Sheraga model \cite{PoSh,Fi},
however taking into account correctly all excluded volume effects.
The melting transition in this model is simply related to the balance 
between the entropic gain of the two DNA strands when being independent,
to be compared with the energy gain in configurations where they 
are tightly bound together.

The numerical results show that excluded volume effects are 
relevant in this transition. They change the transition from being 
second order (without excluded volume effects) to being first order. 
In spite of the latter, we find a divergent length scale, and 
scaling relations typical for a (tri-)critical point. We explain 
this by the absence of a significant surface tension between the molten and 
bound phases. 

The present study seems the first work where excluded volume effects 
are taken into account systematically in DNA melting. But we have 
neglected a number of other features which might be equally 
or even more important. We plan to take some of these into account 
in future work. This includes in particular heterogeneity of the chain
due to the quenched sequence of DNA bases. 

Acknowledgements: We thank Sergio Caracciolo, Walter Nadler, Andrea 
Pelissetto and Lothar Sch{\"a}fer for 
discussions and for carefully reading the manuscript.

\section*{Appendix: Exact results for interacting ideal random walks}

In this appendix we provide an analytic solution for our model in the
ideal case, i.e. when the excluded volume interaction is fully neglected,
on (hyper-)cubic $d$-dimensional lattices.
In this case the generating function can be easily computed and the
order of the transition can be easily determinated in any dimension.
Let us define $c^\eps_{2N}(\x)$ as the unnormalized weight of the configuration of
two $N$-steps random walks $\{\omega^1,\omega^2\}$ starting both at the origin
and ending at a distance vector $\x=\ome^2_N-\ome^1_N$ apart.
A step towards the same lattice position ($\x=0$) is favoured by
an energetic gain $-\eps$, corresponding to a Boltzmann weight
$e^\eps$ (the energy is here expressed in $kT$ units).
We can then write the following recursion relation

\bea
c^\eps_{2N}(\x) &= &\sum_{i=1}^d \sum_{j=1}^d 
   [ c^\eps_{2N-2}(\x+\e_i + \e_j) + c^\eps_{2N-2}(\x+\e_i -\e_j) + 
     c^\eps_{2N-2}(\x-\e_i +\e_j) +\nonumber\\
 &+ &c^\eps_{2N-2}(\x-\e_i-\e_j) ] [1+ ( e^\eps-1 ) \delta_{\x,0}]\;.
\label{rec1}
\eea
Equation \reff{rec1} suggests another interpretation.
We can choose the origin of our system to be solidal at any time step with
the position of one of the two walkers.
In this way the problem is mapped in the one of a single random walker moving at
``double speed" (i.e. making two random steps on the lattice at each time step).
We use here the obvious fact that a walk on a hypercubic lattice can return
to the origin only after an even number of steps (or that the sublattice
with even parity is still a hypercubic lattice). This would not be
true, e.g., for the triangular lattice.

If we denote with $c^\eps_N(\x)$ the weight of the single walker configuration,
the recursion relation \reff{rec1} is equivalent to 
\be
c^\eps_N(\x) = \sum_{i=1}^d [ c^\eps_{N-1}(\x+\e_i) + 
   c^\eps_{N-1}(\x-\e_i) ] [1+ ( e^\eps-1 ) \delta_{\x,0}] \;,
\label{rec2}
\ee
but because of the rescaling of time only even number of steps correspond to the original system.

Introducing the generating function in the grand canonical ensemble by
a Laplace transform
\be
G^\eps(\x,z) = \sum_{N=0}^\infty c^\eps_{N}(\x) z^N,
\ee
one can write
the Fourier transform $\hat G(\q,z)$ in terms of the free propagator
of the Gaussian model on the lattice
\be
D(\q,z)=\frac{1}{m_0^2 + \hat \q^2}\;,
\label{prop}
\ee
where $m_0^2 = \frac{1-2dz}{z}$ and $\hat q_i = 2 \sin \frac{q_i}{2}$, in the following way
\be
\hat G^\eps(\q,z) = \frac{D(\q,z)}{z e^\eps+(1-e^\eps) \int \left[ dq\right]^d D(\q,z)}\;,
\label{master}
\ee
where the integration is done on the first Brillouin zone.
The knowledge of the singular behaviour in $z$ in the small-$q$ region in Eq. \reff{master}, 
makes it possible to determinate the critical
properties of the system in the monodisperse ensemble 
by an inverse Laplace transform.

For instance the partition sum $Z^\eps_{2N} = \sum_\x c^\eps_{2N}(\x)$, the mean value
of powers of components of the square distance {from} the origin (or in the original system of
the square distance of the two walkers) 
$\<x_k^{2m}\> \equiv  \frac{\sum_\x x_k^{2m} c^\eps_{2N}(\x)}{Z^\eps_{2N}}$, 
and the average number of contacts 
$\< n \> = \frac{\partial \log(Z_{2N}^\eps)}{\partial \eps}$
can be obtained by inverse transforms {from} the following
quantities:
\bea
\hat G^\eps(0,z) &=& \sum_\x \sum_{N=0}^{\infty} z^N c^\eps_N(\x) \label{cn} \\
(-1)^m \frac{\partial {\hat G}^\eps(\q,z)}{\partial q_k^{2m}}|_{\q=0} &=&
 \sum_\x \sum_{N=0}^\infty z^N x_k^{2m} c^\eps_N(\x)  \label{x2m} \\
\frac{\partial \hat G^\eps(0,z)}{\partial \eps} &=& 
\sum_\x \sum_{N=0}^{\infty} z^N \frac{\partial c^\eps_N(\x)}{\partial \eps} \label{cont} \\
\eea
Because of the isotropy of the system, we write $\la x_k^{2m}\ra =\la x^{2m}\ra $.
If there were no interaction $(\eps=0)$, the critical behaviour of
the system would arise in the limit $m_0^2 \to 0$, corresponding to $z^* = 1/{2d}$.
We are interested in changes with respect to this critical behaviour due
to the interaction with the origin.

It is immediately clear that a critical behaviour which is different
{from} that of free random walks case can appear in two cases: If the 
denominator in eq. \reff{master} either vanishes or diverges at a $z \le z^*$.

The properties of the denominator are deeply connected to the ones 
of the integral of the free propagator, $H(z)$.

It is easy to see that 
\be
  H(z)=\int \left[ dq\right]^d D(\q,z) \sim \cases{
        (m_0^2) ^{d/2 - 1} &
        for $d < 2$, \cr
  \log m_0^2 &
        for $d = 2$, \cr
   \hbox{\rm finite,} &
        for $d > 2$}
\label{integral}
\ee
so that the case in which the critical behaviour is modified because
of divergencies in the denominator at $z=z^*$ can arise only for
$d \le 2$.

A critical behaviour which characterize a collapsed (double stranded) phase arises when
the relation
\be
\frac{H(z_c(\eps))}{z_c(\eps)}=\frac{1}{1-e^{-\eps}}
\label{betacritico}
\ee
can be satisfied for $z_c < z^*$, while an unstable fixed point 
is reached when Eq. \reff{betacritico} is satisfied for $z_c = z^*$
and the singularity due to the interaction with the origin merges
with the one of the free propagator.

We observe that $H(z)/z$ is an increasing function of $z$
and that $\lim_{z \to 0} H(z)/z =1$,~\footnote{
The assertion follows immediately {from} 
$H(z)/z=\sum_{N=0}^\infty z^N c^{\eps=0}_N(0)$}
for this reason the above equation 
has no solution for $\eps < 0$, the case of repulsive interaction.
But in the region $\eps \ge 0$ one can find a critical value
$\eps^*$ for which Eq. \reff{betacritico} is satisfied for 
$z_c(\eps^*) = z^*$, with $z^*$ defined above. For 
$\eps> \eps^*$ the position of $z_c(\eps)$ in Eq. \reff{betacritico} moves
closer to the origin, and $\lim_{\eps \to \infty} z_c(\eps) =0$.
The point $(\eps^*,z^*)$ is tricritical in the sense 
of having co-dimension $2$ (two control parameters have to be adjusted 
to obtain it).
Eq. \reff{betacritico} gives the critical value $\eps^*$ on the 
cubic lattice which was used in Sec.\reff{id} 
\be
\eps^* =- \log \li 1 -2d H\li\frac{1}{2d} \ri \ri\;.
\label{eps_crit}
\ee

Three kinds of critical behaviour in $z$ can be identified in any dimension
$d$, the one for $(\eps < \eps^*,z=z^*)$ in which the denominator in \reff{master}
does not vanish (molten phase), the one in the {\em double stranded} phase governed by 
$(\eps > \eps^*,z=z_c(\eps))$, and the one at
the tricritical point $(\eps^*,z^*)$, but the behaviour in the different regimes
depends on the dimensionality.

\begin{itemize}

\item{$d \le 2$} 

{From} geometrical arguments it is easy to see that when the dimension of
the space is smaller than the Hausdorff dimension of the walk ($d_H =1/\nu = 2$)
the probability that the walker intersects a given point is finite.
In this case the interaction with the origin results in a 
relevant perturbation that brings the system out of the universality class
of the unperturbed case, as it appears {from} the critical behaviour 
for $m_0^2 \to 0$ of $H(z)$.
The case $d=2$ in which the dimension of the space is equal to the dimension
of the walk corresponds to a marginal interaction and logarithmic corrections appear in
the scaling laws.

Eq. \reff{betacritico} has a solution $z_c(\eps)< z^*$ for all values of $\eps>0$,
and $z_c(\eps) \to 0$ when $\eps \to + \infty$.
This means that in the attractive regime the free propagator stays {\em always} finite when the
denominator vanishes.
In other words
there is {\em no transition between two different regimes when the temperature
is changed if the interaction is attractive}. The system is in the double stranded phase (or
collapsed onto the origin in the other view).
On the other hand we can identify a phase transition when the sign of the interaction changes,
namely at $\eps^* = 0$,
and we will show that it is a second order one.

\item{$d>2$}

In dimension $d>2$ the presence of a zero-dimensional region that couples
with the random walk introduces an irrelevant operator.
The integral $H(z)$ takes a finite value and its expansion around the
singularity $z^*$ of the free propagator has the form \cite{needle}
\be
H(z)= \sum_{n=0}^{\infty} A_n m_0^{2n} + C\left(m_0^2\right) m_0^{d-2},
\qquad m_0^2 = (1-z/z^*)/z\;.
\label{asint}
\ee
where $A_n$ are suitable constants and $C\left(m_0^2\right)$  is a function of
$m_0^2$ which is finite for $m_0^2 \to 0$ for $d$ odd and diverges logarithmically
for $d$ even. 

The leading term in $H(z)-H(z^*)$ for dimension $d<4$ is 
given by the non-analytic one $C\left(m_0^2\right) m_0^{d-2}$, 
it is $m_0^2 \log(m_0^2)$ for $d=4$, and for $d>4$ it is given by the analytic term $m_0^2$. 
We will show that this gives rise to a change in the order of the transition when $d=4$ is crossed.

\end{itemize}

\subsection*{Molten phase:}

In $d>2$, for $\eps < \eps^*$ the critical behaviour in the generating function
is governed by the free propagator $D(\q,z)$ whose behaviour around $z^*$
gives rise to
the usual critical exponents of the random walk, but with amplitudes 
depending on $\eps$.
One can rewrite near to the critical point
\be
\hat G^\eps(\q,z) \approx \frac{D(\q,z)}{z^* e^\eps+(1-e^\eps)H(z^*)}
\label{sopra2}
\ee
and, {from} the small $q$ critical behaviour of $\hat G^\eps(\q,z)$ and of
its derivatives with respect to $q$ and $\eps$, one can obtain the
following scaling forms for large fixed $N$:

partition sum:
\be
Z^\eps_{2N} \approx \frac{z^*}{z^* e^\eps+(1-e^\eps)H(z^*)}{z^*}{^{-2N}}
        \label{cnsopra2} 
\ee
free energy per base pair:
\be
          f =\lim_{N \to \infty} \frac{1}{N}\log Z^\eps_{2N} = \log {{z^*}{^2}}
        \label{fsopra2} 
\ee
base pair separation moment:
\be
\< x^{2m} \> \approx (2m-1)!! {z^*}{^m} 4^m N^m 
        \label{x2msopra2}
\ee
average number of contacts:
\be
\< n \> \approx  \frac{e^\eps \lc H(z^*)-z^* \rc}{z^* e^\eps+(1-e^\eps)H(z^*)}
        \label{nsopra2}
\ee

In the limit $\eps \to {\eps^*}^-$ the mean number of contacts $\< n \>$ and the amplitude 
of $Z^\eps_{2N}$ diverges
\bea
Z^\eps_{2N} &\approx& \frac{z^*}{e^{\eps^*}\lc H(z^*) -z^* \rc } (\eps^* -\eps)^{-1}{z^*}^{-2N}\\
\< n \> &\approx& (\eps^* -\eps)^{-1}\\
\eea
For $d \le 2$, instead, the singular behaviour of $H(z)$ near $z^*$ plays a role
in the molten phase.
For $\eps  < \eps^* =0$, near to the singular region one has
\be
\hat G^\eps (\q,z) \approx \frac{D(\q,z)}{\li 1-\eps \ri H(z)}
\ee
and the above scaling laws become
\bea
Z^\eps_{2N}& \sim & \cases{\li 1-e^{\eps} \ri^{-1}{z^*}^{-2N} N^{d/2 -1} & for $d=1$ \cr
                        \li 1-e^{\eps} \ri^{-1}\lc \log N\rc^{-1}{z^*}^{-2N} & for $d=2$ }\\
f & = &  \log \li {z^*}{^2} \ri \label{bho1}\\
\< n \> & \sim & \frac{1}{e^{-\eps}-1}\\
\< x^{2m} \> & \sim &  N^m\\
\eea
{from} which the behaviour for $\eps \to 0^-$ can be easily recovered.

\subsection*{Double-stranded phase:}

Let us study now the collapsed phase $\eps > \eps^*$.
The denominator determines the critical behaviour, since it vanishes for a $z_c(\eps) < z^*$, 
where the small $q$ behaviour of $D(\q,z)$ is not singular.
Near to the critical region the generating function can be rewritten as
\be
\hat G^\eps(\q,z) \approx \frac{D(\q,z)}{K(\eps) z_c(\eps) \li 1- z/z_c(\eps) \ri}
\label{sotto2}
\ee
where $K(\eps) =(e^{\eps}-1) \frac{\partial H}{\partial z}|_{z_c(\eps)} - e^\eps$ is positive.
The derivative with respect to $\eps$ of \reff{master} in the same limit reads
\be
\frac{\partial \hat G^\eps(\q,z)}{\partial \eps} \approx
\frac{e^\eps \li H(z_c(\eps))-z_c(\eps) \ri}{z_c(\eps)^2 K^2 (\eps) \li 1-z/z_c(\eps) \ri^2}D(\q,z)
\ee
In the same way as the previous relations for the observables have been recovered,
one finds that 
\bea
Z^\eps_{2N} & \approx & \frac{z_c(\eps)^{-2N}}{K(\eps)\li 1-z_c(\eps)/z^* \ri}\label{cnsotto2}\\
f &=& \log\li {z_c(\eps)}{^2}\ri \label{fsotto2}\\
\< x^{2m} \>& \approx  & (2m-1)!! m! \frac{z_c(\eps)^m}{\li 1-z_c(\eps)/z^* \ri^m} \label{x2msotto2}\\
\< n \>& \approx  & \frac{e^\eps \lc H(z_c(\eps))-z_c(\eps) \rc}{z_c(\eps) K(\eps) }N\label{nsotto2}
\eea

It is interesting to study the limit $\eps \to {\eps^*}^{+}$, whose comparison
with $\eps \to {\eps^*}^{-}$ gives the order of the transition.
The behaviour of the denominator near to $\eps^*$ and the way in which $z_c(\eps)$
approaches $\eps^*$ will be crucial. 

We have to distinguish different cases with respect to the dimensionality
of the space.

Let us start with $d>2$. {From} Eq. \reff{integral} it follows that 
\be
  z e^{\eps^*}+\li 1-e^{\eps^*}\ri H(z) \sim \li 1-e^{\eps^*}\ri \cases{
        \li 1-z/z^* \ri^{d/2-1} &
        for $2<d<4$, \cr
        \li 1-z/z^* \ri \log \li 1-z/z^* \ri &
        for $d=4$, \cr
        \li 1-z/z^* \ri &
        for $d > 4$}
\label{chisei}
\ee
{From} this relation it is a simple matter to get the behaviour of $z^*-z_c (\eps)$
and $K(\eps)$ for $\eps \simg \eps^*$
\bea
z^*-z_c(\eps) &\sim& \cases{
        \li \eps - \eps^* \ri^{2\over{d-2}} & for $2<d<4$, \cr
        \li \eps - \eps^* \ri \li \log\li \eps - \eps^* \ri \ri^{-1}& for $d=4$, \cr
        \li \eps - \eps^* \ri & for $d>4$.}
\label{betaeps}\\
K(\eps)& \sim& \cases{
        \li \eps - \eps^* \ri^{{d-4}\over{d-2}} & for $2<d<4$, \cr
        \log\li \eps - \eps^* \ri & for $d=4$, \cr
        \hbox{constant} & for $d>4$.}
\eea
The shift exponent $\psi$, defined by 
\be
z_c(\eps) = z^* \li 1 - k \li \eps -\eps^* \ri^{1/\psi} \ri\;,
\ee
with $k$ constant,
can be read {from} the above relations \reff{betaeps} as a function of $d$.
Introducing the crossover exponent $\phi$ 
\be
\phi=\cases{ \frac{d-2}{2} & for $2<d \le 4$, \cr
		1 & for $d > 4$}\;,
\ee
the above relations can be rewritten in a more compact form, namely
$z^*-z_c(\eps)\sim\li \eps - \eps^* \ri^{1/\phi}$, $K(\eps)\sim \li \eps - \eps^* \ri^
\frac{\phi-1}{\phi}$, both with the appropriate logarithmic correction in $d=4$.
One can observe that $\phi$ coincides with the shift exponent $\psi$.
It follows that for $\eps \simg  {\eps^*}$, 
\bea
Z^\eps_{2N} &\approx& \frac{z_c(\eps)^{-2N}}{\eps -\eps^*}\\
f &\approx& \log\li {z^*}{^2}\ri + a \cases{ \li \eps - \eps^* \ri^{1/\phi} & for $d \ne 4$ \cr
                                        \li \eps - \eps^* \ri \lc \log \li \eps - \eps^* \ri \rc^{-1} & for $d=4$}\;,\\
\< x^{2m} \>& \approx& (2m-1)!! m! \cases{ {\li \eps - \eps^* \ri^{-m/\phi}} &  for $d \ne 4$ \cr
                     { \lc \log \li \eps - \eps^* \ri \rc^{m}}{\li \eps - \eps^* \ri^{-m}}& for $d=4$}\;,\label{termica}\\
\< n \>& \approx& e^{\eps^*} \lc \frac{H(z^*)}{\beta^*}-1\rc N \cases{
 \li \eps - \eps^* \ri^\frac{1-\phi}{\phi}  & for $d \ne 4$ \cr 
 \lc \log \li \eps - \eps^* \ri \rc^{-1} & for $d=4$}\;.
\eea    
{From} \reff{asint} one sees immediately that the free energy has a 
discontinuity in its first derivative if $d>4$, i.e. the transition is a first order one.

On the other hand a different critical behaviour is expected at the critical point
$\eps =\eps^*$.
In this special point the denominator vanishes exactly at $z_c(\eps)=z^*$,
the point in which the free propagator has the small $q$ singularity.
The behaviour of the denominator of $\hat G^{\eps^*}(\q,z)$  , its derivatives with
respect to $q$
and $ \partial_{\eps}\hat G^{\eps}(\q,z) |_{\eps^*}$
for $z$ near to $z^*$ can
be read {from} the asymptotic expansion in \reff{chisei}.

Using the same exponent $\phi$ defined above, the following scaling laws hold

\bea
Z^{\eps^*}_{2N} & \sim & \cases{ {z^*}^{-2N} N^\phi & for $d\ne 4$ \cr
                                {z^*}^{-2N} {N^\phi}\lc {\log N}\rc^{-1} & for $d=4$ }\;,\label{tr_id}\\
\< x^{2m} \>& \sim & \frac{(2m-1)!! m! \Gamma(1+\phi){z^*}^{m} }{\Gamma(m+1+\phi)} 4^m N^m \;,\\
\< n \>& \sim & \cases{ N^\phi & for $d\ne 4$ \cr
                                {N^\phi}\lc {\log N}\rc^{-1} & for $d=4$ }\;.
\eea

The case $d \le 2$ can be handled in similar way knowing the behaviour of $H(z)$ 
near to $z^*$. 
The way in which $z_c(\eps)$ approaches $z^*$ and $K(\eps)$ diverges
when $\eps \to 0$ is given by
\bea
\li z^* - z_c(\eps) \ri &\sim& \cases{\eps^\frac{2}{2-d}  & for $d=1$ \cr
                                         e^{-a/\eps} & for $d=2$ }\;,\\
K(\eps)& \sim& \cases{\eps^{-\frac{2}{2-d}} & for $d=1$ \cr
                               e^{a/\eps} \eps & for $d=2$ }\;,
\eea
with $a$ constant.
It follows that the above observables in the limit $\eps \to 0^+$ behave as
\bea
Z^{\eps}_{2N} & \sim & \cases{ {z^*}^{-2N}  & for $d=1$ \cr
                        {z^*}^{-2N} \eps^{-1}& for $d=2$   }\\
f & \approx & \log \li {z^*}{^2}\ri +c \cases{\eps^2 & for $d=1$ \cr
                                         e^{-a/\eps} & for $d=2$ }\label{bho}\\
\< n \>& \approx& N\cases{\eps & for $d=1$ \cr
                                         e^{-a/\eps} \eps^{-2}& for $d=2$ }
\eea
Taking the limit $\eps \to 0^+$ in Eq. \reff{bho} and $\eps \to 0^-$ in Eq. \reff{bho1}
it  is a simple matter to see that the transition is a second order one.

It's a simple matter to see what is the behaviour at the tricritical point.
$\hat G^0(\q,z)$ becomes the free generating function and the scaling laws that
appear are the usual ones.
The number of contacts can be simply recovered {from} the derivative of \reff{master}
with respect to $\eps$ setting $\eps =0$.
This gives the well known result
\be
\< n \> \sim \cases{N^{1/2} & for $d=1$\cr  
                        \log N & for $d=2$}
\ee

\end{document}